\documentclass[a4paper,fleqn,usenatbib]{mnras}

\usepackage{amsmath,graphicx,amssymb}
\usepackage{txfonts}
\usepackage{natbib}
\usepackage{subfigure}
\usepackage{hyperref}
\usepackage{adjustbox}

\def\logTeff{\ensuremath{\log T_{\mathrm{eff}}}}
\def\logl{\ensuremath{\log L/L_{\odot}}}
\defcitealias{lb16}{LB16}

\title[High-Amplitude $\gamma$ Doradus Variables]{High-Amplitude $\gamma$ Doradus Variables}

\author[Paunzen et al.]{
Ernst Paunzen$^{1}$\thanks{E-mail: epaunzen@physics.muni.cz}
Klaus Bernhard,$^{2,3}$
Stefan H{\"u}mmerich,$^{2,3}$
Franz-Josef Hambsch,$^{2,3,4}$
\newauthor
Christopher Lloyd,$^{5}$
Sebasti\'an Otero$^{3}$
\\
$^{1}$Department of Theoretical Physics and Astrophysics, Masaryk University, Kotl\'a\v{r}sk\'a 2, 611 37 Brno, Czech Republic\\
$^{2}$Bundesdeutsche Arbeitsgemeinschaft f{\"u}r Ver{\"a}nderliche Sterne e.V. (BAV), Berlin, Germany\\
$^{3}$American Association of Variable Star Observers (AAVSO), Cambridge, USA\\
$^{4}$Vereniging Voor Sterrenkunde (VVS), Brugge, BE-8000, Belgium\\
$^{5}$Department of Physics and Astronomy, University of Sussex, Falmer, Brighton BN1 9QH, UK\\
}

\date{Accepted 2020 September 18. Received 2020 September 18; in original form 2020 April 16}

\pubyear{2020}

\begin{document}

\label{firstpage}
\pagerange{\pageref{firstpage}--\pageref{lastpage}}
\maketitle

\begin{abstract} %currently 224 words
According to most literature sources, the amplitude of the pulsational variability observed in $\gamma$ Doradus stars does not exceed 0.1\,mag in Johnson $V$. We have analyzed fifteen high-amplitude $\gamma$ Doradus stars with photometric peak-to-peak amplitudes well beyond this limit, with the aim of unraveling the mechanisms behind the observed high amplitudes and investigating whether these objects are in any way physically distinct from their low-amplitude counterparts. We have calculated astrophysical parameters and investigated the location of the high-amplitude $\gamma$ Doradus stars and a control sample of fifteen low-amplitude objects in the \logTeff\ versus \logl\ diagram. Employing survey data and our own observations, we analyzed the photometric variability of our target stars using discrete Fourier transform. Correlations between the observed primary frequencies, amplitudes and other parameters like effective temperature and luminosity were investigated. The unusually high amplitudes of the high-amplitude $\gamma$ Doradus stars can be explained by the superposition of several base frequencies in interaction with their combination and overtone frequencies. Although the maximum amplitude of the primary frequencies does not exceed an amplitude of 0.1\,mag, total light variability amplitudes of over 0.3\,mag ($V$) can be attained in this way. Low- and high-amplitude $\gamma$ Doradus stars do not appear to be physically distinct in any other respect than their total variability amplitudes but merely represent two ends of the same, uniform group of variables.
\end{abstract}


%%%%%%%%%%%%%%%%%%%%%%%%%%%%%%%%%%%%%%%%%%%%%%%%%%

%%%%%%%%%%%%%%%%% BODY OF PAPER %%%%%%%%%%%%%%%%%%

\section{Introduction} \label{introduction}

The $\gamma$ Doradus and $\delta$ Scuti stars are pulsating variables that are situated in the region of the A and F-type main sequence stars. For convenience, they are referred to hereafter as, respectively, GDOR and DSCT stars, according to their designations in the General Catalogue of Variable Stars (GCVS; \citealt{gcvs}). In contrast to the long-known and very well studied DSCT stars \citep{fath35,breger00}, the variability of GDOR stars was discovered relatively recently. They were identified as a new class of variables by \citet{balona94} and defined as such by \citet{kaye99}.

The GDOR stars are characterized by high-order, low-degree, non-radial gravity (g) mode pulsation \citep{kaye99}, which is thought to be driven by the convective flux blocking mechanism \citep{guzik00,dupret05}. They are encountered between spectral types A7 and F7 (GCVS), although other sources have shifted the red border of the GDOR instability strip to somewhat hotter temperatures. \citet{balona11}, for instance, find most GDOR stars in the effective temperature range from 6500\,$<$\,$T_{\mathrm{eff}}$\,$<$7000\,K, corresponding to spectral types F1 to F5 on the main sequence, while \citet{bradley15} find all of their GDOR candidates between 6100\,$<$\,$T_{\mathrm{eff}}$\,$<$7500\,K. However, it has been shown that there are also hot GDOR stars, which are located between the red edge of the Slowly Pulsating B star and the blue edge of the GDOR star instability strips \citep{balona16,ali20}.

GDOR and DSCT stars can be distinguished by the timescales of the observed variability, although the instability strips for both classes overlap and hybrid-types exist \citep[e.g.][]{henry05}. Several different classification systems are found in the literature. According to the GCVS, GDOR stars exhibit variability in the period range of 0.3\,$\leq$\,$P$\,$\leq$\,3\,d (0.33\,$\leq$\,$f$\,$\leq$\,3.33\,d$^{-1}$) (cf. \citealt{kaye99}), while DSCT stars are encountered in the period range of 0.01\,$\leq$\,$P$\,$\leq$\,0.2\,d (5\,$\leq$\,$f$\,$\leq$\,100\,d$^{-1}$). Based on an analysis of high-precision $Kepler$ photometry, \citet{grigahcene10} proposed a division into 'pure' DSCT stars ($f$\,$>$\,5\,c/d), 'pure' GDOR stars ($f$\,$<$\,5\,c/d), and hybrid-types exhibiting variability in both frequency regimes. While hybrid-type pulsators are mostly discovered in ultra-precise space photometry, their frequency of occurrence ($\sim$25\% of the sample of \citealt{uytterhoeven11}) has made clear that the situation is complex and the traditional classification scheme might be in need of revision. The understanding and relationship of DSCT and GDOR pulsators is currently in flux \citep[e.g.][]{grigahcene10,uytterhoeven11,balona11,balona12a,balona18,antoci19}.

Using $Kepler$ data, \citet{balona11} described three groups of distinct GDOR star light curves. Two of these groups show pronounced beating effects. These are the SYM stars, which show more or less symmetric light curves, and the ASYM stars, whose light curves are asymmetric in the sense that the beat amplitude is larger when the star is brighter, which results in large variations in maximum brightness but only small variations in minimum brightness. The third group is made up of the MULT stars, which are characterized by many low-amplitude peaks that do not lead to pronounced beating in the light curve.

\citet{kaye99} indicate a photometric peak-to-peak amplitude\footnote{In this paper, unless indicated otherwise, amplitude always refers to peak-to-peak amplitude.} of up to 0.1\,mag in Johnson $V$ for GDOR variables. This limit has been widely accepted and is found throughout the literature and variability catalogues, like e.g. the International Variable Star Index of the AAVSO (VSX; \citealt{vsx}). The GCVS somewhat softens this statement, indicating that peak-to-peak amplitudes are \textit{usually} up to 0.1\,mag. In fact, at the time of this writing (March 2020), only 24 out of the 924 GDOR variables contained in the VSX are listed with amplitudes exceeding 0.1 mag, and the knowledge on these objects is limited.

This paper presents a detailed investigation of fifteen high-amplitude ($V$\textsubscript{amp}$\,>\,$0.1\,mag) GDOR (referred to hereafter for convenience as HAGDOR = \textbf{H}igh-\textbf{A}mplitude \textbf{G}amma \textbf{DOR}adus) stars with survey data and our own observations, with the aim of unraveling the mechanisms behind the observed high amplitudes and investigating possible systematic differences between the group of the regular ($V$\textsubscript{amp}$\,\le\,$0.1\,mag) GDOR stars and the HAGDORs.

\begin{figure}
\begin{center}
\includegraphics[width=0.47\textwidth]{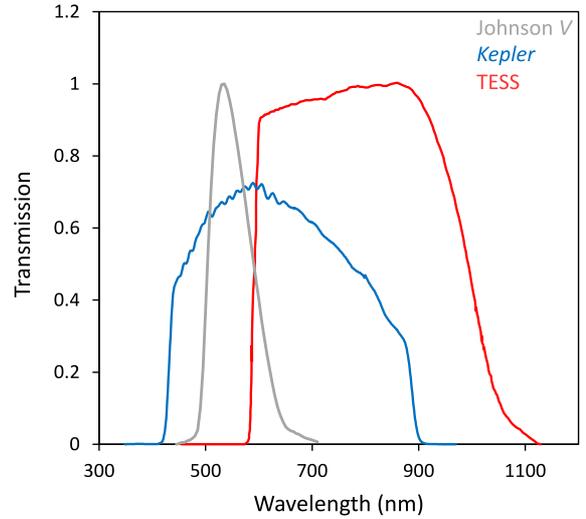}
\caption{Response functions of the Johnson $V$, $Kepler$ and TESS passbands. Data have been gleaned from \citet{Johnson51}, \citet{TESS3} and the $Kepler$ Instrument Handbook (https://keplerscience.arc.nasa.gov/).}
\label{response_functions}
\end{center}
\end{figure} 

%%%%%%%%%%%%%%%%%%%%%%%%%%%%%%%%%%%%%%%%%%%%%%%%%%%%%%%%%%%%%

\section{Target stars and astrophysical parameters} \label{targets}

\subsection{Target selection} \label{target_selection}

The VSX was chosen as first source for selecting our initial sample stars as it is the most current and accurate variable star database available. At the time of this writing, it listed a total of 24 GDOR stars out of 924 stars with amplitudes exceeding 0.1\,mag -- including KIC\,8113425, which was analyzed in detail by \citet{kurtz15}. Ten of these objects boasted suitable photometric time series data allowing further detailed analyses and were hence selected for our sample. The stars KIC\,7448050, KIC\,6953103 and KIC\,7304385, erroneously listed in the VSX with amplitudes less than 0.1\,mag, were subsequently identified as HAGDOR stars and also included into our sample. As these examples illustrate, we suspect that, on detailed analysis, more low-amplitude GDOR stars listed in the VSX will likely turn out to be HAGDOR stars. Amplitude determination in these objects, which often show pronounced beating effects in their light curves, is not easy and prone to errors. An investigation into this matter, however, is beyond the scope of the present paper. Finally, two more HAGDORs (HD\,33575 and HD\,211394) were identified in a systematic search among unclassified variables of suitable spectral type in the VSX. Both objects exhibit very large peak-to-peak amplitudes of more than 0.3\,mag ($V$). In total, our final HAGDOR sample consists of 15 stars showing light change amplitudes in excess of 0.1\,mag ($V$). Table \ref{table1} provides essential data for these objects.

For the six HAGDOR stars having both broadband $Kepler$ ($Kp$) data and $V$-band data from the All-Sky Automated Survey for Supernovae (ASAS-SN; cf. Section \ref{targets}), we calculated and compared semi-amplitudes in the different passbands and found that Amp($V$)\,/\,Amp($Kp$)\,=\,1.08(5). Assuming a colour-amplitude ratio of $\sim$1.25 for GDOR stars \citep{handler02} and employing the relations for the calibration of $Kp$ magnitudes given by \citet{brown11}, we estimate Amp($V$)\,/\,Amp($Kp$)\,=\,1.065(20), in line with the above mentioned result. We have therefore adopted Amp($V$)\,/\,Amp($Kp$)\,=\,1.07(3) for the purposes of the present study.

Reduced peak-to-peak amplitudes in $Kepler$ data are expected, as photometric pulsation amplitudes in early-type stars generally decrease with increasing wavelength, and the $Kepler$ passband covers the wavelength range from 420-900\,nm, with peak transmission at around 600\,nm (cf. Section \ref{Kepler}). As only one sample star boasts TESS data (HD\,17721), a similar estimation of the relationship between Amp($V$) and Amp($TESS$) was not possible. However, we expect that pulsation amplitudes will be even more reduced in the redder TESS passband (600-1000\,nm; cf. Section \ref{TESS}). This is in line with the results of \citet{antoci19}, who investigated DSCT and GDOR stars with TESS data and estimated that pulsation amplitudes derived from TESS data only reach 74(1)\,\% of those derived from $Kepler$ data. Using Amp($V$)\,/\,Amp($Kp$)\,=\,1.07, we estimate Amp($V$)\,/\,Amp($TESS$)\,=\,1.44(4). The response functions of the Johnson $V$, $Kepler$ and TESS passbands are shown in Fig. \ref{response_functions}.

\begin{figure}
\begin{center}
\includegraphics[width=0.47\textwidth]{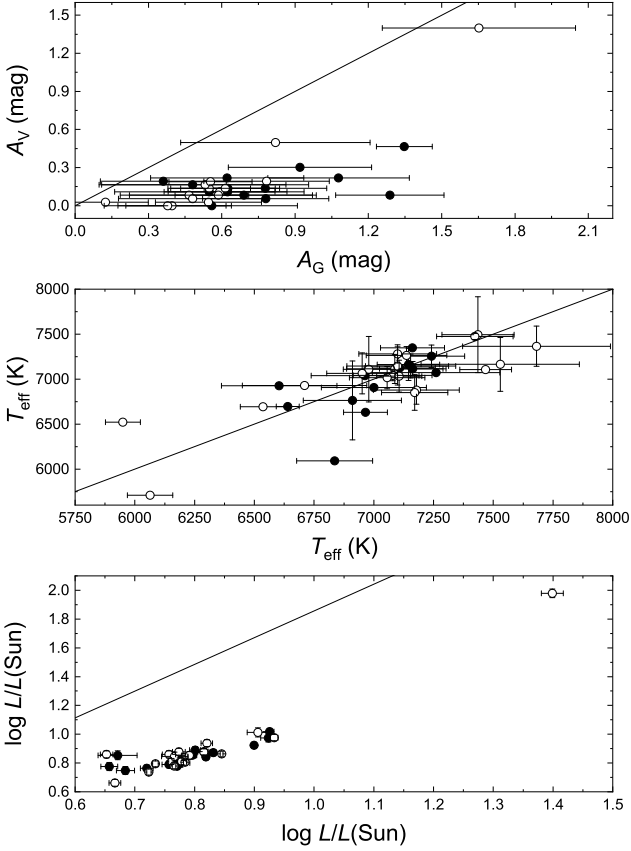}
\caption{Comparison of the reddening (upper panel), effective temperature (middle panel), and luminosity (lower panel) values derived in this paper (ordinate values; Table \ref{table_parameters}) and from GAIA DR2 (abscissa values). Filled and open circles denote HAGDOR and GDOR stars, respectively. Also indicated are the unity lines.}
\label{fig_comp_para}
\end{center}
\end{figure} 

\begin{figure}
\begin{center}
\includegraphics[width=0.47\textwidth]{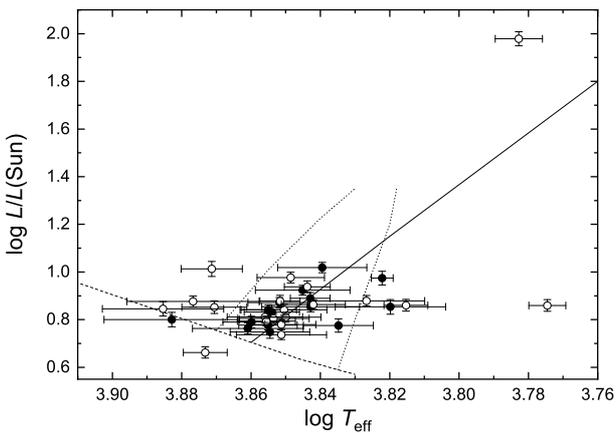}
\caption{The \logTeff\ versus \logl\ diagram of our programme stars (Table \ref{table_parameters}). The red border of the DSCT (solid line), and the GDOR (dotted lines) instability strips are taken from \citet{breger98} and \citet{dupret04}, respectively. The zero-age main sequence (dashed line) is taken from \citet{claret95}. Filled and open circles denote HAGDOR stars and GDOR stars, respectively.}
\label{fig_instability_strip}
\end{center}
\end{figure}

To investigate systematic differences between the groups of HAGDOR and GDOR variables, a control sample of regular GDOR stars was selected from the VSX using the following criteria: (1) light change amplitudes of 0.05\,$\le$$V$$\le\,$0.1\,mag, (b) the availability of high-precision photometric time series data of suitable length that allow an in-depth analysis, (c) the availability of reliable astrophysical parameters, and (d) similar effective temperatures to the HAGDOR stars. The lower amplitude limit (1) was chosen because of the limitations of the employed ground-based photometric time series data (cf. also Section \ref{targets}). Most GDOR stars in VSX have listed amplitudes that fail to satisfy criterion (1); therefore only 20 stars were selected. The stars of the regular GDOR sample can be gleaned from the lower part of Table \ref{table_parameters}.

\begin{table*}
\caption{Essential data for the sample HAGDOR stars, listed in order of increasing right ascension. The columns denote: (1) conventional identifier; (2) alternative identification; (3) right ascension (J2000); (4) declination (J2000); (5) peak-to-peak amplitude; the corresponding data source is provided in parentheses (A=ASAS-SN/A3=ASAS-3/K=$Kepler$/R=ROAD/T=TESS); (6) main period (d); (7) most recent spectral type from the literature. Positional information was taken from Data Release 2 (DR2) of the Gaia satellite mission \citep{gaia1,gaia2,gaia3}. Information on the relationship of the amplitudes in the different passbands can be gleaned from Section \ref{target_selection}.}
\label{table1}
\begin{center}
\begin{adjustbox}{max width=\textwidth}
\begin{tabular}{lllllll}
\hline
\hline																																																																																
(1) & (2) & (3) & (4) & (5) & (6) & (7) \\
Object	 & 	Alt. ID	 & 	RA\,(J2000)	 & 	Dec\,(J2000)	 & 	Amp.	(Source) & 	Main per. (d)	& Spec. Type (lit) \\
\hline
GSC\,02831-00348 & V0758\,And	& 02 22 20.479 & $+$37 59 05.107 & 0.19 (A) & 0.924655	& n/a \\
HD\,17721	& HIP\,13089 & 02 48 15.087	& $-$57 39 42.498	& 0.14 (A3), 0.144 (T)	& 1.087858 & A9 V \citep{1975mcts.book.....H} \\
HD\,33575 & NSV\,1858 & 05 10 21.600 & $-$23 13 01.950 & 0.35 (A3), 0.35 (A) & 1.293260	& A9 V \citep{1988mcts.book.....H}	\\
HD\,50875	& NSV\,3272	& 06 54 33.947 & $-$11 23 29.468 & 0.22 (A3) & 1.705329	& F2 V \citep{2003AJ....125..359W}	\\
HD\,85693 & NSV\,18291 & 09 52 56.364	& $-$26 45 20.219	& 0.17 (A3)	& 1.125952 & F0 V \citep{1982mcts.book.....H} \\
GSC\,09046-00646 & ASAS\,J163451-6446.3	& 16 34 50.969 & $-$64 46 18.299 & 0.21 (A3), 0.19 (A) & 1.180647 & n/a \\
HD\,150538 & NSV\,20738 & 16 41 29.100 & $+$06 16 33.387 & 0.21 (A3) & 1.386903	& F5 \citep{1980BICDS..19...74O} \\
KIC\,3847822 & TYC\,3134-2121-1 & 19 22 57.126 & $+$38 58 18.167 & 0.16 (A), 0.169 (K) & 1.204964 & n/a \\
KIC\,3441414 & GSC\,03134-00901	& 19 23 21.727 & $+$38 32 58.614 & 0.10 (A), 0.122 (K) & 0.810700	& F0 V \citep{2016AJ....151...13G}	\\
KIC\,7448050 & ASAS\,J193103+4302.1	& 19 31 03.390 & $+$43 02 06.368 & 0.19 (A), 0.182 (K) & 0.877616	& A9 IV-V \citep{2016AJ....151...13G} \\
KIC\,6953103 & 2MASS\,J19325124+4228465 & 19 32 51.238 & $+$42 28 46.505 & 0.24 (A), 0.249 (K) & 0.776640 & F0 V (this work)$^{1}$ \\
KIC\,8113425 & 2MASS\,J19474808+4354257	& 19 47 48.082 & $+$43 54 25.727 & 0.14 (A), 0.164 (K) & 2.325268	& F0 V \citep{frasca16} \\
KIC\,7304385 & ASAS\,J195052+4248.1 & 19 50 51.541 & $+$42 48 06.015 & 0.14 (A), 0.146 (K) & 0.787874 & F5 \citep{Skiff} \\
HD\,211394 & BD-17\,6481 & 22 16 57.007 & $-$16 59 03.156 & 0.32 (A3), 0.29 (R), 0.324 (K) & 2.210397 & F0 V \citep{2014ASPC..485..223B} \\					
GSC\,02780-02174 & TYC\,2780-2174-1 & 23 51 25.338 & $+$37 10 27.924 & 0.13 (A) & 0.943619 & F0 V (this work)$^{1}$ \\							

\hline
\hline
\multicolumn{7}{l}{$^{1}$ Derived from analysis of a publicly available LAMOST DR4 spectrum (\citealt{lamost1,lamost2}).} \\
\end{tabular}                                                                                                                                                             
\end{adjustbox}
\end{center}                                                                                                                                    
\end{table*}

%%%%%%%%%%%%%%%%%%%%%%%%%%%%%%%%%%%%%%%%%%%%%%%%%%%%%%%%%%%%%

\subsection{Astrophysical parameters} \label{astrophysical_parameters}

Gaia DR2 \citep{gaia1,gaia2,gaia3} includes effective temperatures, luminosities, and reddening values for nearly all our sample stars. As a first step, we checked the reliability of these parameters.

Effective temperatures were gleaned from the literature \citep{ammons06,mcdonald12,pinsonneault12,huber14,munari14,decat15,frasca16,kunder17}, and mean values and standard errors were calculated. Unfortunately, for one HAGDOR star (GSC\,02831-00348), no parameters are available in the above-listed references. Reddening values were interpolated using the maps published by \citet{Green2018}. Distances and their errors were derived from Gaia DR2 parallax data. Almost all our sample stars are located within 1\,kpc from the Sun. Therefore, reddening is small but not negligible ($A_{\mathrm{V}}$\,$<$\,0.2\,mag). The only exception is the GDOR star GSC\,04281-00186, which is located in the Galactic disk ($l$\,$=$\,$-$0.45\degr).

To calculate luminosities of our sample stars, bolometric corrections (B.C.) and relative magnitudes $V$ were needed. B.C. values are at a minimum for F-type stars \citep{Pecaut13} and do not significantly influence the luminosity calculation. Unfortunately, no homogeneous source of $V$ magnitudes is available for all our target stars. Therefore, mean values of the magnitudes published by \citet{kharchenko01} and \citet{Henden16} were calculated and $G$ magnitudes from Gaia DR2 were transformed. The final mean astrophysical parameters are listed in Table \ref{table_parameters}.

Figure \ref{fig_comp_para} shows a comparison between the values derived in this paper and the values derived from Gaia DR2. It becomes obvious that the Gaia reddening values are significantly larger than those derived from the reddening maps (upper panel). As a consequence, the corresponding luminosities are lower than the luminosities calibrated with the parallax data and the other observables (lower panel). The situation for the effective temperatures (middle panel) is different. With only three exceptions, all stars are located on the unity line within the errors. For each of the three outliers, only one effective temperature value is available in the literature. Therefore, it is not possible to investigate the reason for, and estimate the significance of, the outlying positions in the diagram. In consequence, for the following analyses, we have employed our own luminosity values and effective temperatures from Gaia DR2 because the latter have been derived in a homogeneous way.

Figure \ref{fig_instability_strip} presents the \logTeff\ versus \logl\ diagram of our sample stars. Also shown are the GDOR \citep{dupret04} instability strip and the red border of the DSCT \citep{breger98} instability strip. No obvious differences are seen between the location of the GDOR and HAGDOR stars; both groups are well distributed over the whole main sequence up to \logl\,$<$\,1.1. Interestingly, most stars are also located in the DSCT instability strip but do not show any corresponding pulsations with a detectable amplitude in the here employed photometric time series data.

Two objects deserve mention, which are situated well outside the instability strips. These are the GDOR stars GSC\,04281-00186 and GSC\,09289-02186. According to its calibrated astrophysical parameters, GSC\,09289-02186 is a G0\,V star. GSC\,04281-00186 is apparently a G-type giant situated in a significantly reddened ($A_{\mathrm{V}}$\,=\,1.4\,mag) region at a Galactic latitude of $b$\,$\approx$\,0\,$^{\circ}$. Nevertheless, even when neglecting reddening, we find the star still far above the terminal-age main sequence. Using the standard reddening correlation $A(V)$\,=\,3.45$A(J)$\,=\,5.89$A(H)$\,=\,7.69$A(K_{\mathrm{s}}$) \citep{2017BlgAJ..26...45P}, the derived indices $(J-H)_0$\,=\,$-$0.115\,mag and $(H-K_{\mathrm{s}})_0$\,=\,+0.040\,mag are typical for an early B-type star \citep{2009BaltA..18...19S}. However, Gaia DR2 colours (and all others in the optical region) are typical for a G-type object. Also, the derived effective temperature from Gaia DR2 is in agreement with all other published values. The star's status as an evolved object, therefore, seems to be beyond doubt. This is intriguing as giant stars are not expected to exhibit GDOR pulsation. Both objects are further discussed in Section \ref{comparison}.

\begin{table*}
\caption{Mean astrophysical parameters of our target stars. The upper part of the table contains the HAGDOR stars, the lower part the GDOR stars. The columns denote: (1) conventional identifier; (2) alternative identification; (3,4) absorption in $V$ and $G$; (5) bolometric correction, (6) $V$ magnitude and error; (7,8) logarithmic effective temperature and error estimate; (9,10) logarithmic luminosity and error estimate.}
\label{table_parameters}
\begin{center}
\begin{adjustbox}{max width=\textwidth}
\begin{tabular}{llllllllll}
\hline
\hline
(1) & (2) & (3) & (4) & (5) & (6) & (7) & (8) & (9) & 10 \\
Object & Alt. ID & $A_{\mathrm{V}}$ & $A_{\mathrm{G}}$ & B.C. & $V$ & \logTeff & \logTeff & \logl & \logl \\
& & (our) & (DR2) & (our) & (our) & (our) & (DR2) & (our) & (DR2) \\
\hline
GSC\,02831-00348	&  V0758\,And	        		&	0.11	&	0.55(39)	&	+0.00	&	11.390(4)	&		$-$	&	3.883(20)	&	0.80(3)	&	0.78(2)	\\                 
HD\,17721		&  HIP\,13089 				&	0.00	&	0.56(35)	&	$-$0.01	&	8.598(1)	&	3.852(5)	&	3.855(8)	&	0.78(2)	&	0.77(1)	\\                 
HD\,33575		&  NSV\,1858 				&	0.05	&	0.78(26)	&	$-$0.01	&	9.708(2)	&	3.866		&	3.855(8)	&	0.84(2)	&	0.82(1)	\\                 
HD\,50875		&  NSV\,3272				&	0.08	&	0.69(28)	&	$-$0.01	&	8.521(2)	&	3.853		&	3.852(9)	&	0.87(2)	&	0.83(1)	\\                 
HD\,85693		&  NSV\,18291 				&	0.03	&		$-$	&	$-$0.01	&	7.679(2)	&	3.839(1)	&	3.845(14)	&	0.92(2)	&	0.90(1)	\\                 
GSC\,09046-00646 	&  ASAS\,J163451-6446.3			&	0.22	&	0.62(31)	&	$-$0.01	&	10.251(2)	&	3.822		&	3.843(6)	&	0.89(2)	&	0.80(1)	\\                 
HD\,150538		&  NSV\,20738 				&	0.22	&	1.08(29)	&	$-$0.02	&	9.731(2)	&	3.830(28)	&	3.839(13)	&	1.02(2)	&	0.93(1)	\\                 
KIC\,3847822		&  TYC\,3134-2121-1 			&	0.14	&	0.63(19)	&	$-$0.01	&	11.866(1)	&	3.850(1)	&	3.861(16)	&	0.76(2)	&	0.72(1)	\\                 
KIC\,3441414		&  GSC\,03134-00901			&	0.16	&	0.48(38)	&	$-$0.01	&	11.539(2)	&	3.855(11)	&	3.854(8)	&	0.83(3)	&	0.77(1)	\\                 
KIC\,7448050		&  ASAS\,J193103+4302.1			&	0.11	&	0.62(24)	&	$-$0.01	&	11.838(2)	&	3.861(14)	&	3.860(8)	&	0.79(2)	&	0.76(1)	\\                 
KIC\,6953103		&  2MASS\,J19325124+4228465		&	0.19	&	0.36(26)	&	$-$0.01	&	12.593(2)	&	3.854(11)	&	3.855(11)	&	0.75(3)	&	0.68(2)	\\                 
KIC\,8113425		&  2MASS\,J19474808+4354257		&	0.47	&	1.35(11)	&	$-$0.04	&	13.931(1)	&	3.841(8)	&	3.820(16)	&	0.85(3)	&	0.67(3)	\\                 
KIC\,7304385		&  ASAS\,J195052+4248.1 		&	0.14	&	0.78(25)	&	$-$0.01	&	10.078(1)	&	3.847(18)	&	3.843(14)	&	0.85(2)	&	0.80(1)	\\                 
HD\,211394		&  BD-17\,6481 				&	0.08	&	1.29(22)	&	$-$0.04	&	9.306(3)	&	3.826(9)	&	3.822(3)	&	0.97(3)	&	0.92(1)	\\                 
GSC\,02780-02174	&  TYC\,2780-2174-1 			&	0.30	&	0.92(29)	&	$-$0.02	&	11.754(2)	&	3.785		&	3.835(10)	&	0.78(3)	&	0.66(1)	\\                 
\hline
HD\,18011 		&  HIP\,13494              		&	0.14	&	0.55(17)	&	$-$0.01	&	9.201(1)	&	3.852(10)	&	3.851(12)	&	0.84(2)	&	0.78(1)	\\ %HIP 13494      
CD-87\,32		&  TYC\,9500-1039-1        		&	0.13	&	$-$		&	$-$0.03	&	10.283(1)	&	3.841		&	3.827(17)	&	0.88(2)	&	0.82(1)	\\ %CSTAR 070941   
BD-12\,1502 		&  NSV\,16873              		&	0.11	&	0.59(23)	&	$-$0.01	&	8.937(1)	&	3.848(4)	&	3.850(10)	&	0.81(2)	&	0.76(1)	\\ %NSV 16873      
EPIC\,202072613 & TYC\,1342-1962-1    &    0.03    &    0.12(18)    &    0.00    &    11.276(1)    &    3.852        &    3.873(6)    &    0.66(2)    &    0.67(1)    \\
CD-23\,9345		&  TYC\,6620-698-1         		&	0.08	&	0.47(24)	&	$-$0.01	&	10.018(2)	&	3.837(10)	&	3.856(11)	&	0.81(2)	&	0.78(1)	\\ %ASAS J103117-23
HD\,124248 & MU\,Vir &    0.00    &    0.38    0.26    &    -0.01    &    7.156(2)    &    3.862(6)    &    3.851(10)    &    0.78(2)    &    0.77(1)    \\
HD\,135825 &   IN\,Lib  &    0.00    &    0.40    0.22    &    -0.01    &    7.285(2)    &    3.854(13)    &    3.851(13)    &    0.74(2)    &    0.72(1)    \\
GSC\,08298-00090	&  TYC\,8298-90-1          		&	0.50	&	0.82(39)	&	$-$0.04	&	11.214(2)	&	3.826		&	3.815(6)	&	0.86(2)	&	0.65(1)	\\ %ASAS J152138-47
GSC\,09289-02186	&  V0366\,Aps              		&	0.20	&	$-$		&	$-$0.09	&	11.522(1)	&	3.814		&	3.775(5)	&	0.86(2)	&	0.76(1)	\\ %V0366 Aps      
HD\,164615 & V2118\,Oph    &    0.00    & $-$ &    -0.01    &    6.995(1)    &    3.849(14)    &    3.842(9)    &    0.86(2)    &    0.85(1)    \\
KIC\,12643786		&  TYC\,3554-1916-1        		&	0.08	&	0.59(40)	&	+0.00	&	11.470(1)	&	3.855(18)	&	3.877(19)	&	0.88(2)	&	0.82(1)	\\                 
KIC\,11080103		&  2MASS\,J19185013+4837138		&	0.14	&	0.61(22)	&	+0.00	&	12.875(1)	&	3.867(13)	&	3.885(17)	&	0.85(3)	&	0.77(2)	\\                 
KIC\,5105754		&  TYC\,3139-2577-1        		&	0.16	&	0.53(42)	&	$-$0.01	&	11.355(1)	&	3.847(11)	&	3.852(6)	&	0.88(2)	&	0.77(1)	\\                 
KIC\,4757184		&  TYC\,3139-499-1         		&	0.19	&	$-$		&	+0.00	&	11.777(1)	&	3.875(24)	&	3.871(9)	&	1.01(3)	&	0.91(2)	\\                 
KIC\,11920505		&  TYC\,3564-2927-1        		&	0.05	&	0.48(30)	&	$-$0.01	&	9.772(1)	&	3.836(13)	&	3.856(8)	&	0.79(2)	&	0.73(1)	\\                 
KIC\,11826272		&  BD+49\,3115             		&	0.19	&	0.78(26)	&	$-$0.01	&	10.294(2)	&	3.852(22)	&	3.844(7)	&	0.94(2)	&	0.82(1)	\\                 
HD\,187615 & V1844\,Aql &    0.03    &    0.55    0.22    &    -0.01    &    7.939(1)    &    3.861(6)    &    3.854(10)    &    0.80(2)    &    0.78(1)    \\ 
GSC\,09086-01560 	&  TYC\,9086-1560-1        		&	0.19	&	0.55(17)	&	+0.00	&	11.229(2)	&	3.874(2)	&	3.871(9)	&	0.85(3)	&	0.79(1)	\\ %Romanov V9     
HD\,218225 		&  DI Gru                 		&	0.09	&	$-$	&	$-$0.01	&	8.724(1)	&	3.846(8)	&	3.849(10)	&	0.98(2)	&	0.93(1)	\\ %DI Gru         
GSC\,04281-00186	&  TYC\,4281-186-1         		&	1.40	&	1.65(40)	&	$-$0.08	&	11.276(1)	&	3.757		&	3.783(7)	&	1.98(3)	&	1.40(2)	\\ %TYC 4281-186-1 
\hline
\hline
\end{tabular}                                                                                                                                                             
\end{adjustbox}
\end{center}                                                                                                                                    
\end{table*}
	
%%%%%%%%%%%%%%%%%%%%%%%%%%%%%%%%%%%%%%%%%%%%%%%%%%%%%%%%%%%%%

\section{Data sources and method of analysis} \label{targets}

\subsection{The ASAS-3 photometric archive} \label{ASAS3}

Phase 3 of the All Sky Automated Survey (ASAS-3) lasted from 2000 until 2009 \citep{pojmanski02} and monitored the entire southern sky and part of the northern sky ($\delta$\,$<$\,+28\degr). The ASAS-3 system was situated at the 10-inch astrograph dome of the Las Campanas Observatory in Chile and boasted two wide-field telescopes equipped with f/2.8 200\,mm Minolta lenses and 2048 x 2048 AP 10 Apogee detectors. About 10${^7}$ sources were monitored in the Johnson $V$ passband. The ASAS-3 archive contains photometry for stars in the magnitude range 7\,$\la$\,$V$\,$\la$\,14; the most accurate photometry was obtained for stars in the range of 8\,$\la$\,$V$\,$\la$\,10, boasting a typical scatter of about 0.01\,mag \citep{pigulski14}.

The long time baseline of almost 10 years renders the detection of periodic signals with very small amplitudes possible. For instance, \citet{david14} identified periodic variables with a peak-to-peak amplitude of 0.01 -- 0.02\,mag in the magnitude range of 7\,$\la$\,$V$\,$\la$\,10. According to \citet{pigulski14}, the detection of periodic signals in the frequency range of 0\,$<$\,$f$(d$^{-1}$)\,$<$\,40 with amplitudes as low as about 5\,millimag (mmag) is possible.

%%%%%%%%%%%%%%%%%%%%%%%%%%%%%%%%%%%%%%%%%%%%%%%%%%%%%%%%%%%%%

\subsection{The ASAS-SN photometric archive} \label{ASASSN}

The ASAS-SN survey is monitoring the entire visible sky every night to a depth of $V$\,$\la$\,17 mag \citep{shappee14,kochanek17}. The available data span up to six years of observations. As of end-2017, ASAS-SN observations are procured at five stations, each consisting of four 14\,cm aperture Nikon telephoto lenses. Observations consist of three dithered 90\,s exposures made through $V$ (two stations) or $g$ (three stations) band filters. ASAS-SN saturates at 10 to 11\,mag, where the exact limit depends on the camera and the image position (vignetting). However, a procedure inherited from the original ASAS survey is applied which corrects for saturation but increases the noise in the affected data sets \citep{jayasinghe18}.

%%%%%%%%%%%%%%%%%%%%%%%%%%%%%%%%%%%%%%%%%%%%%%%%%%%%%%%%%%%%%

\subsection{The $Kepler$ satellite photometric archive} \label{Kepler}

The $Kepler$ satellite was launched in March 2009, with the primary goal of detecting transiting exoplanets in the solar neighbourhood. The spacebased photometer has a 0.95\,m aperture; the detectors consist of 21 modules each equipped with two 2200x1024 pixel CCDs. $Kepler$ provides single passband (420-900\,nm; \citealt{Koch10}) light curves of micromagnitude precision, taken in long-cadence (29.5\,min) and short-cadence (58.5\,s) modes \citep{Gilliland10}, and has discovered hundreds of planet candidates \citep{Kepler}. The long, uninterrupted, and high-precision time series photometry is ideally suited to the study of multiperiodic pulsating stars \citep{tkachenko12}.

%%%%%%%%%%%%%%%%%%%%%%%%%%%%%%%%%%%%%%%%%%%%%%%%%%%%%%%%%%%%%

\subsection{The TESS satellite photometric archive} \label{TESS}

The Transiting Exoplanet Survey Satellite (TESS) mission is a two-year all-sky survey aiming at the discovery of transiting exoplanets \citep{TESS1}. To this end, four MIT/Lincoln Lab CCDs with 4096x4096 pixels are employed (imaging area of 2048x2048 pixels; the remaining pixels are used as a frame-store to allow rapid shutterless readout). The cameras have an effective aperture size of 10\,cm and are equipped with f/1.4 lenses, resulting in a field of view of 24\degr\,x\,24\degr\ per camera. The TESS passband covers the wavelength range from about 600-1000\,nm. Due to their high photometric precision, time sampling of 2\,min and long intervals of uninterrupted observations, TESS data are well  suited to asteroseismology \citep{TESS2}.

%%%%%%%%%%%%%%%%%%%%%%%%%%%%%%%%%%%%%%%%%%%%%%%%%%%%%%%%%%%%%

\subsection{The Remote Observatory Atacama Desert (ROAD)} \label{ROAD}

New CCD photometric observations of one target (HD 211394) were acquired at the Remote Observatory Atacama Desert (ROAD; \citealt{ROAD}). All observations were acquired through an Astrodon Photometric $V$ filter with an Orion Optics, UK Optimized Dall Kirkham 406/6.8 telescope and a FLI 16803 CCD camera. The exposure time was 5\,s; twilight sky-flat images were used for flatfield corrections. Reductions were performed with the MAXIM DL program\footnote{http://www.cyanogen.com}. For the determination of magnitudes, the LesvePhotometry program\footnote{http://www.dppobservatory.net/} was employed. 

%%%%%%%%%%%%%%%%%%%%%%%%%%%%%%%%%%%%%%%%%%%%%%%%%%%%%%%%%%%%%

\subsection{Method of analysis} \label{data_analysis}

The data of our target stars were downloaded from the ASAS-3 website\footnote{http://www.astrouw.edu.pl/asas/}, the ASAS-SN archive\footnote{https://asas-sn.osu.edu/} and, in the case of $Kepler$ and TESS data, the Mikulski Archive for Space Telescopes (MAST).\footnote{https://archive.stsci.edu/access-mast-data} All light curves were inspected visually. Obvious outliers and data points with very large uncertainties were carefully removed. In the case of the ASAS-3 data, measurements with a quality flag of 'D' (= 'worst data, probably useless') were deleted.

The time of the ASAS-3 and ASAS-SN observations are provided in HJD-2450000. $Kepler$ and TESS data, however, are formatted as BJD-2454833 and BJD-2457000, respectively. To facilitate analysis, $Kepler$ and TESS data have been converted to HJD-2450000 to bring them in line with the ground-based data.\footnote{In all cases, the calculation of the phase values provided in the presentation of results has been based on the time basis of HJD-2450000.}

The period analysis was done using the program package \textsc{PERIOD04} \citep{period04}, which employs discrete Fourier transform and allows least-squares fitting of multiple frequencies to the data. To extract all relevant frequencies, the data were searched for periodic signals and consecutively prewhitened with the most significant frequencies. As detection threshold, we adopted S/N\,$\ge$\,4 \citep{breger93}.

\section{Presentation and discussion of results} \label{discussion}

\begin{figure}
\begin{center}
 \mbox{
		\parbox{0.47\textwidth}{
		\subfigure{\includegraphics[width=0.47\textwidth]{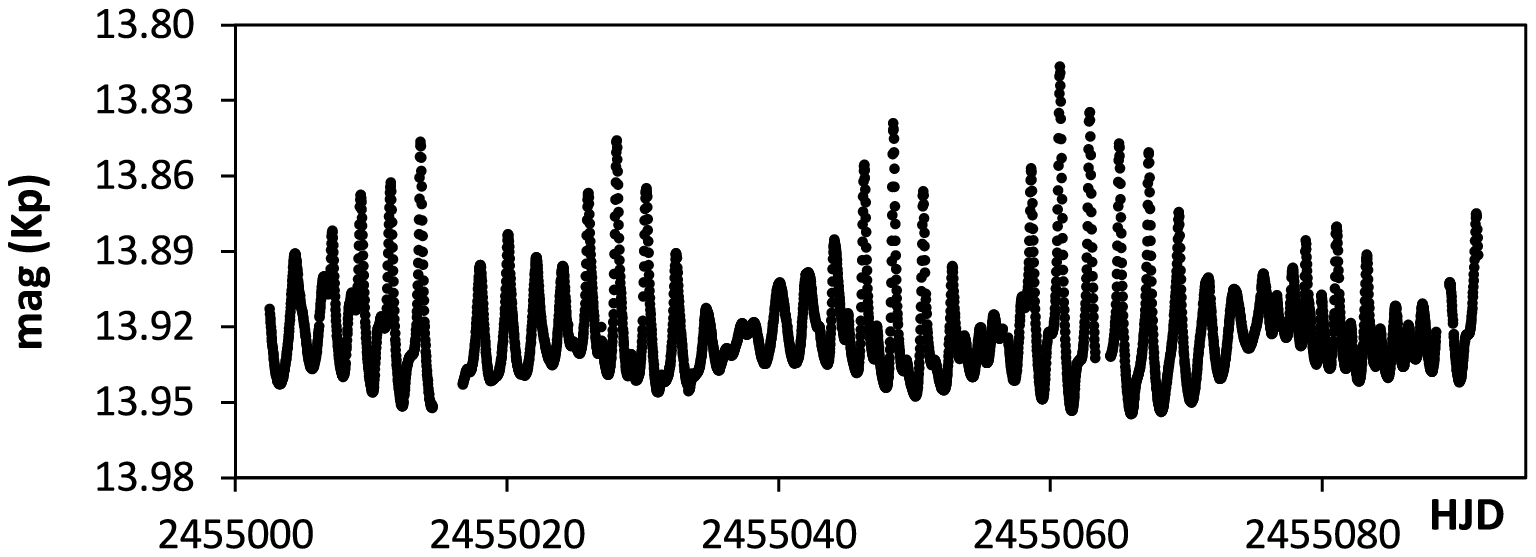}}
		\subfigure{\includegraphics[width=0.47\textwidth]{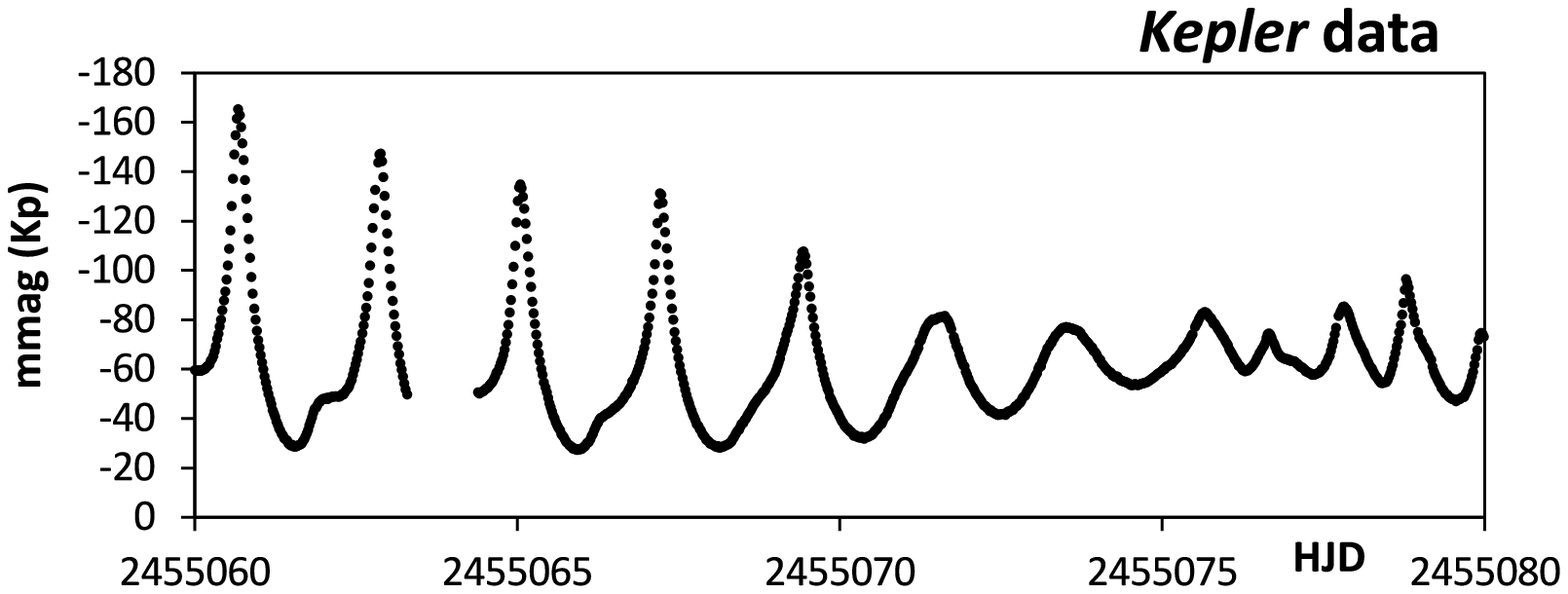}}
		\subfigure{\includegraphics[width=0.47\textwidth]{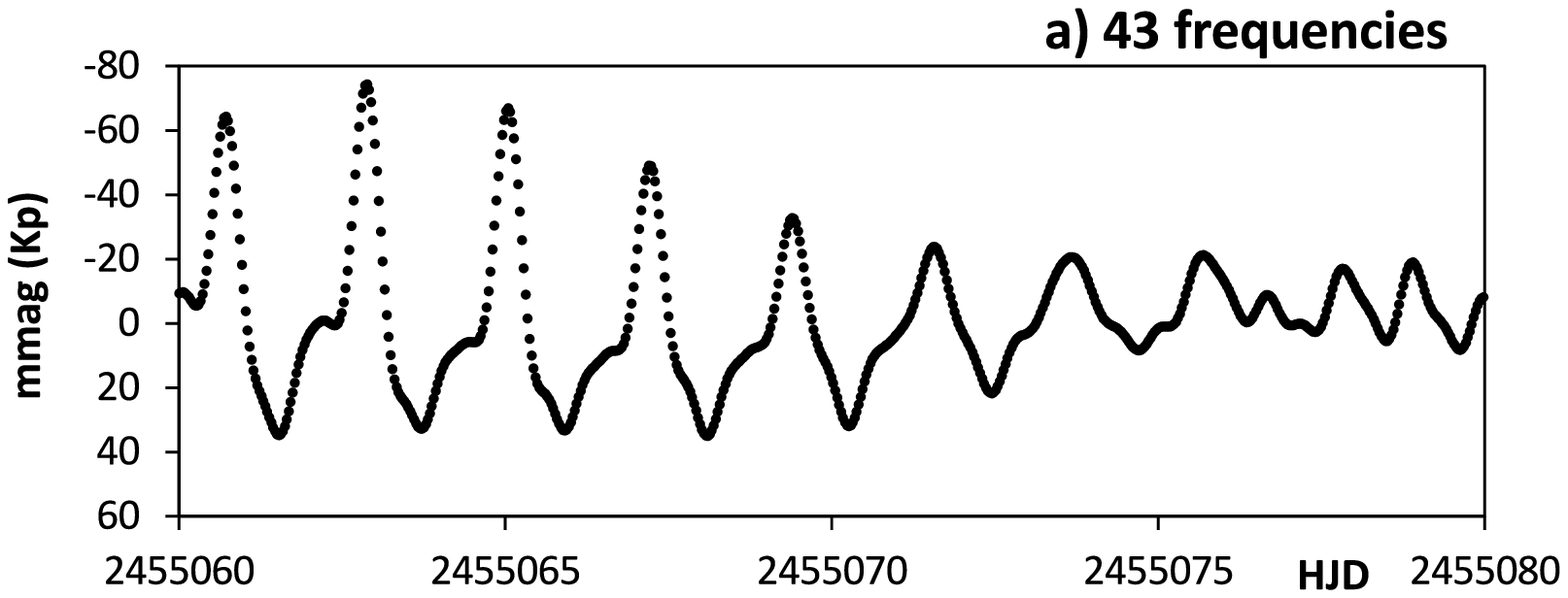}}
		\subfigure{\includegraphics[width=0.47\textwidth]{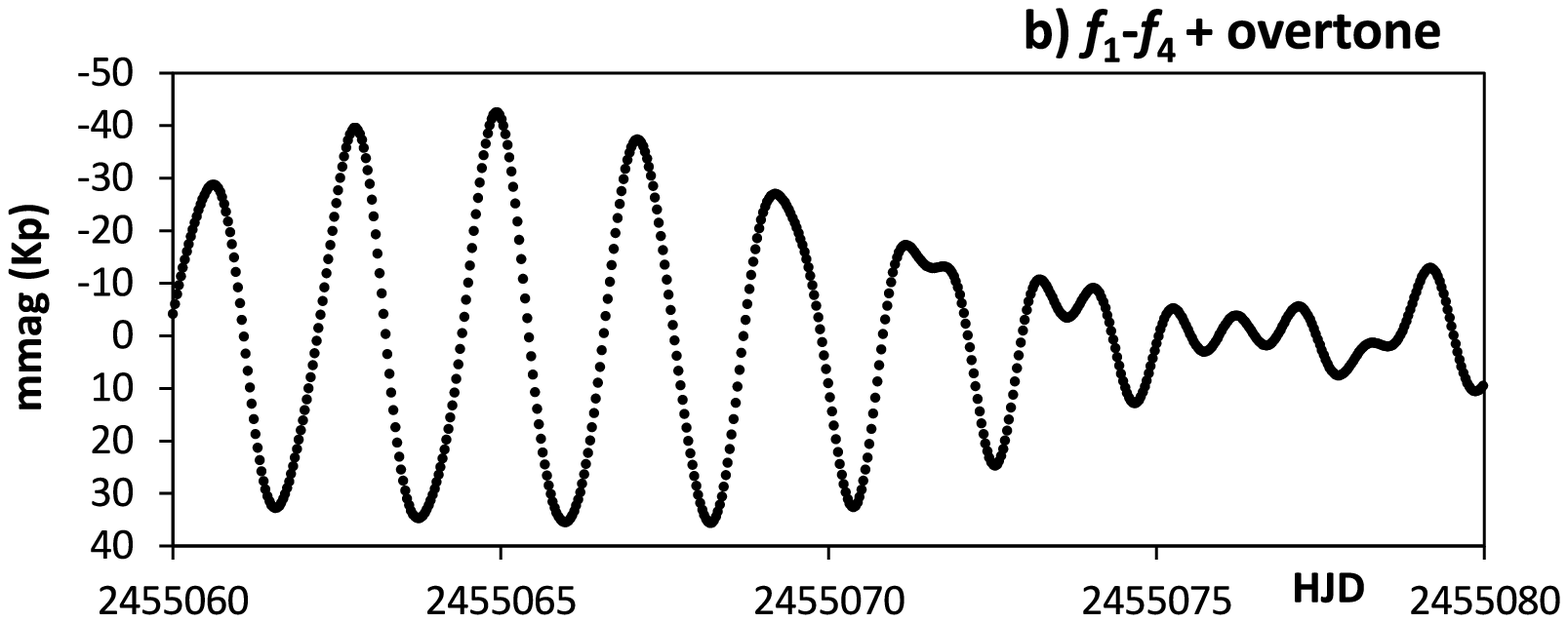}}
		\subfigure{\includegraphics[width=0.47\textwidth]{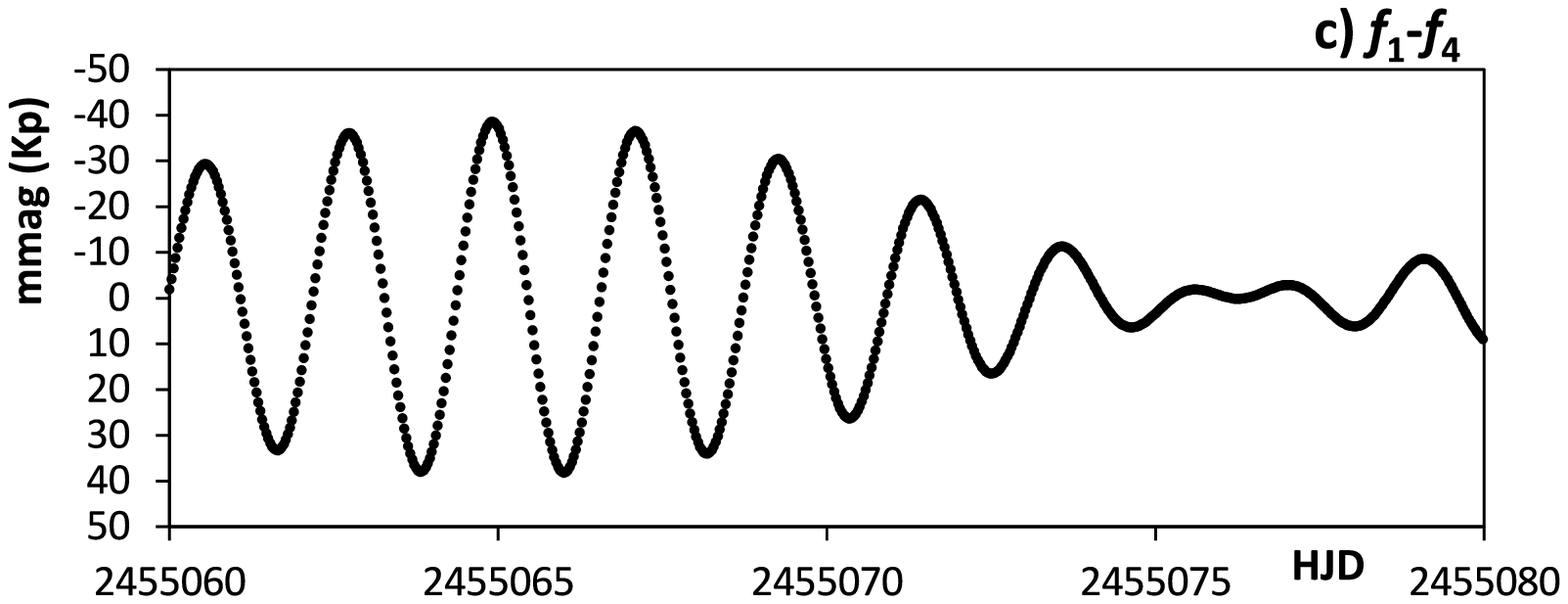}}
    \subfigure{\includegraphics[width=0.47\textwidth]{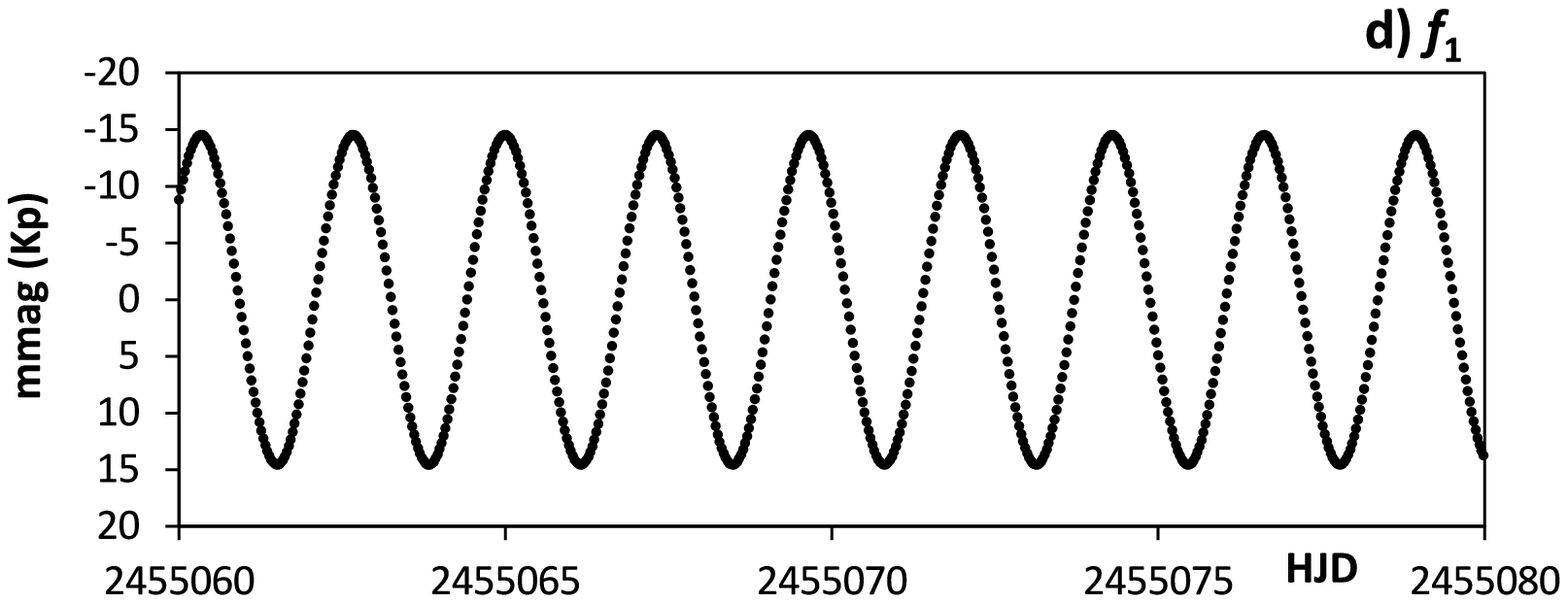}}}}
\caption{The two top panels show the full $Kepler$ light curve of KIC\,8113425 and the light curve in the interval HJD\,2455060-80. The other panels illustrate the simulated light curves using a) 43 frequencies, b) seven frequencies, c) four frequencies and d) one frequency, as described in \citet{kurtz15}.}
\label{fig1}
\end{center}
\end{figure}

In this section, we present example results and discuss the light variability patterns of the HAGDOR variables on the basis of the HAGDOR star KIC\,8113425, which has been extensively studied by \citet{kurtz15} and is here employed as a model case.

\subsection{The HAGDOR star KIC\,8113425 as a model case}

In their investigation of the complex frequency spectra of GDOR, Slowly Pulsating B stars and Be stars, \citet{kurtz15} extensively studied the HAGDOR star KIC\,8113425. It has therefore been employed as a model case for a general interpretation of the light variations of our sample HAGDOR stars which also include KIC\,8113425. \citet{kurtz15} noted the complex, strongly non-linear light variations of this star, which shows a much larger range at maximum light than minimum light (type ASYM), and identified 43 frequencies with semi-amplitudes greater than 1\,mmag that cluster in five frequency groups. All these 43 frequencies can be understood in terms of only four base frequencies ($f_{1}$\,=\,0.430058\,d$^{-1}$, $f_{2}$\,=\,0.450101\,d$^{-1}$, $f_{3}$\,=\,0.461264\,d$^{-1}$ and $f_{4}$\,=\,0.489414\,d$^{-1}$) and their combination frequencies up to the order 2$f$ (e.g. 2$f_{1}$, $f_{2}$\,+\,$f_{3}$\,$-$\,$f_{4}$). To trace and investigate the results of \citet{kurtz15}, simulated light curves were calculated based on the frequency and phase information provided by the aforementioned authors.

The full $Kepler$ light curve of KIC\,8113425 and a detailed view are shown in the top panels of Figure \ref{fig1}. The other panels of this figure illustrate the simulated light curves using
\begin{enumerate}
	\item[a)] all 43 frequencies,
	\item[b)] seven frequencies (four base frequencies and the three detected overtone frequencies 2$f_{1}$, 2$f_{3}$ and $f_{4}$),
	\item[c)] four frequencies (the base frequencies), and
	\item[d)] one frequency ($f_{1}$, the frequency with the largest semi-amplitude, 0.01455\,mag).
\end{enumerate}

We have investigated the contribution of the corresponding frequencies to the total variability amplitude. The base frequency $f_{1}$ with its peak-to-peak amplitude of 0.028\,mag accounts for only $\sim$21\,\% of the total amplitude (0.137\,mag) of the observed light variations. The four base frequencies together add up to an amplitude of about 0.076\,mag ($\sim$55\,\% of the total amplitude), while the four base frequencies plus the three overtone frequencies reach an amplitude of 0.0858\,mag ($\sim$62\,\%). The 43 frequencies add up to an amplitude of 0.109\,mag, which is $\sim$80\,\% of the total amplitude. According to the conclusions of \citet{kurtz15}, suitable combination frequencies can result in considerably larger peak-to-peak amplitudes than the base frequency alone, as has been shown above. This is in agreement with the finding that, apart from these special cases, GDOR variables usually show low amplitudes.

%%%%%%%%%%%%%%%%%%%%%%%%%%%%%%%%%%%%%%%%%%%%%%%%%%%%%%%%%%%%%

\begin{figure*}
\begin{center}
	\includegraphics[width=\textwidth]{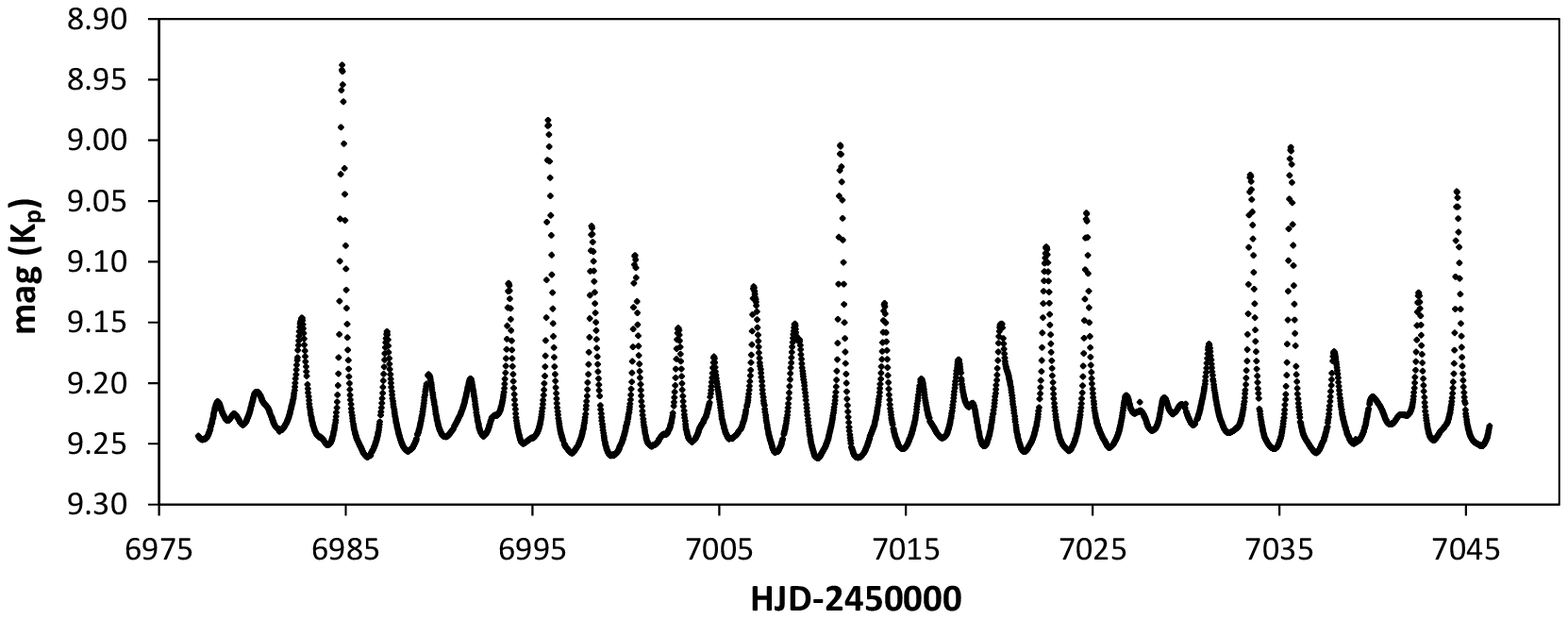}
	\includegraphics[width=\textwidth]{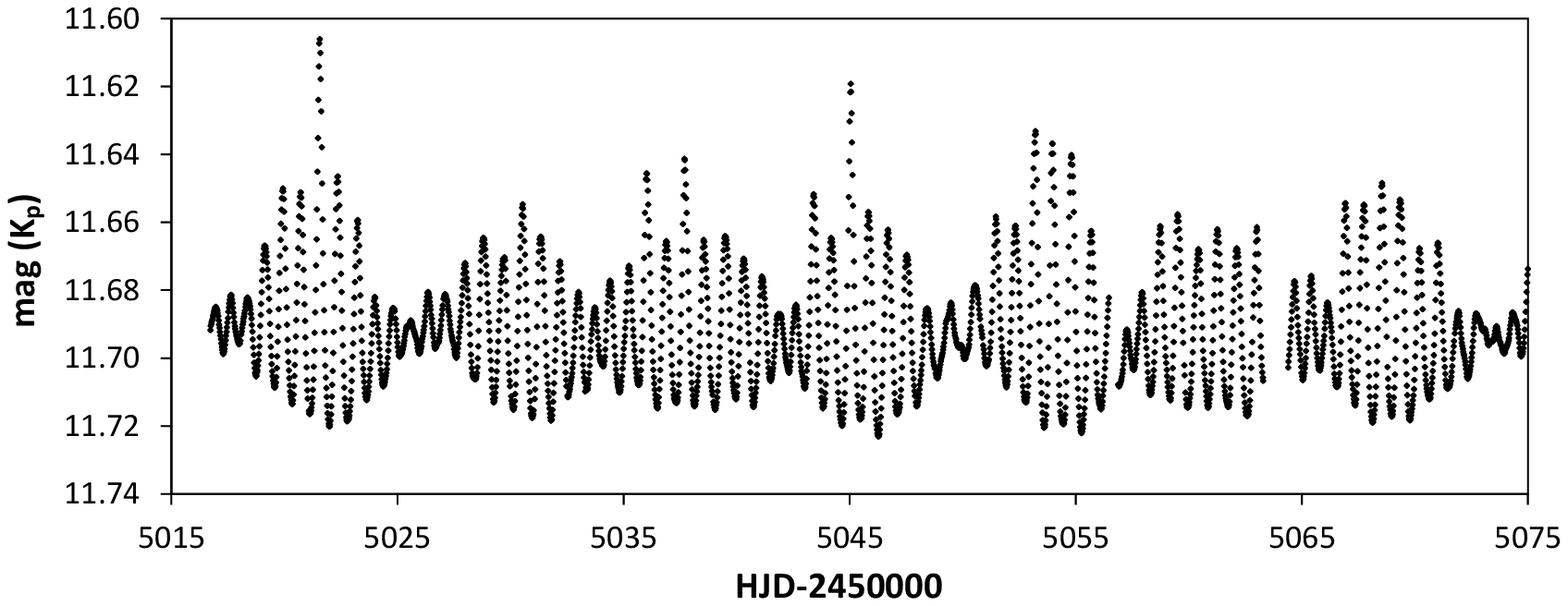}
\caption{Detailed view of parts of the $Kepler$ light curves of HD\,211394 (upper panel) and KIC\,3441414 (lower panel), highlighting the beating effects in the light curves and the high peak-to-peak amplitudes of 0.324 and 0.122\,mag, respectively. Both stars show light curves of the ASYM type \citep{balona11}.}
\label{fig_light_curves1}
\end{center}
\end{figure*}

\begin{table}
\caption{Frequency solution for \textbf{HD\,211394}.}
\label{fs_example1}
\begin{center}
\begin{adjustbox}{max width=0.5\textwidth}
\begin{tabular}{lllll}
\hline
\hline
(1) & (2) & (3) & (4) & (5) \\
Freq.No. & Frequency & Amplitude & Phase & ID \\
\hline
$F1\textsubscript{ASAS-3}$ & 0.452407	& 0.049	& 0.7275 & $f_{1}$ \\
$F2\textsubscript{ASAS-3}$ & 0.371387	& 0.027	& 0.6153 & $f_{2}$ \\
$F3\textsubscript{ASAS-3}$ & 0.904804	& 0.020	& 0.7111 & $2f_{1}$ \\
$F4\textsubscript{ASAS-3}$ & 0.823776	& 0.018	& 0.6548 & $f_{1}+f_{2}$ \\
$F5\textsubscript{ASAS-3}$ & 1.276179	& 0.014	& 0.6848 & $f_{2}+2f_{1}$  \\
\hline
$F1\textsubscript{ROAD}$ & 0.452625 & 0.051 & 0.0744 & $f_{1}$ \\
$F2\textsubscript{ROAD}$ & 0.370860 & 0.026 & 0.7140 & $f_{2}$ \\
$F3\textsubscript{ROAD}$ & 0.823041 & 0.020 & 0.4374 & $f_{1}+f_{2}$ \\
$F4\textsubscript{ROAD}$ & 0.904372 & 0.018 & 0.1150 & $2f_{1}$ \\
\hline
$F1\textsubscript{Kepler}$ & 0.452538 & 0.04193 & 0.8082 & $f_{1}$ \\
$F2\textsubscript{Kepler}$ & 0.372089 & 0.01975 & 0.7546 & $f_{2}$ \\
$F3\textsubscript{Kepler}$ & 0.905160 & 0.01774 & 0.2723 & $2f_{1}$ \\
$F4\textsubscript{Kepler}$ & 0.824686 & 0.01274 & 0.4111 & $f_{1}+f_{2}$ \\	
$F5\textsubscript{Kepler}$ & 0.080533 & 0.01063 & 0.3164 & $f_{1}-f_{2}$ \\
$F6\textsubscript{Kepler}$ & 1.005225 & 0.01002 & 0.3856 & $f_{3}$ \\
$F7\textsubscript{Kepler}$ & 0.533652 & 0.00904 & 0.2196 & $2f_{1}-f_{2}$ \\
$F8\textsubscript{Kepler}$ & 1.357718 & 0.00816 & 0.2133 & $f_{4}$ \\
\hline
\end{tabular}
\end{adjustbox}
\end{center}  
\end{table}  

\begin{table}
\caption{Frequency solution for \textbf{KIC\,3441414}.}
\label{fs_example2}
\begin{center}
\begin{adjustbox}{max width=0.5\textwidth}
\begin{tabular}{lllll}
\hline
\hline
(1) & (2) & (3) & (4) & (5) \\
Freq.No. & Frequency & Amplitude & Phase & ID \\
\hline
$F1\textsubscript{ASAS-SN}$ & 1.233502	& 0.021	& 0.5729 & $f_{1}$ \\
$F2\textsubscript{ASAS-SN}$ & 1.107843	& 0.010	& 0.7617 & $f_{2}$ \\
$F3\textsubscript{ASAS-SN}$ & 1.321001	& 0.007	& 0.2928 & $f_{3}$ \\
\hline
$F1\textsubscript{Kepler}$ & 1.233479	& 0.01841	& 0.1633 & $f_{1}$ \\	
$F2\textsubscript{Kepler}$ & 1.107875	& 0.01237	& 0.8904 & $f_{2}$ \\
$F3\textsubscript{Kepler}$ & 1.320999	& 0.00622	& 0.6805 & $f_{3}$ \\
$F4\textsubscript{Kepler}$ & 0.125600 & 0.00350	& 0.1636 & $f_{1}-f_{2}$ \\
$F5\textsubscript{Kepler}$ & 2.341354	& 0.00230	& 0.3444 & $f_{1}+f_{2}$ \\
$F6\textsubscript{Kepler}$ & 0.543345	& 0.00176	& 0.3657 & $f_{4}$ \\	
$F7\textsubscript{Kepler}$ & 0.962388	& 0.00170	& 0.4148 & $f_{5}$ \\
$F8\textsubscript{Kepler}$ & 2.466959	& 0.00164 & 0.6123 & $2f_{1}$	\\
$F9\textsubscript{Kepler}$ & 1.465986	& 0.00138	& 0.7528 & $f_{6}$ \\
$F10\textsubscript{Kepler}$ & 0.087513	& 0.00134	& 0.4919 & $f_{3}-f_{1}$ \\
$F11\textsubscript{Kepler}$ & 1.179790	& 0.00130	& 0.4088 & $f_{7}$ \\	
$F12\textsubscript{Kepler}$ & 0.213140	& 0.00120	& 0.6362 & $f_{3}-f_{2}$ \\
$F13\textsubscript{Kepler}$ & 0.508685	& 0.00109	& 0.6989 & $f_{8}$ \\
$F14\textsubscript{Kepler}$ & 1.661927	& 0.00109	& 0.3922 & $f_{9}$ \\
$F15\textsubscript{Kepler}$ & 1.742302	& 0.00108	& 0.0020 & $f_{10}$ \\
$F16\textsubscript{Kepler}$ & 1.359081	& 0.00105	& 0.6474 & $2f_{1}-f_{2}$ \\
$F17\textsubscript{Kepler}$ & 0.034023	& 0.00102	& 0.4082 & $-2f_{1}+f_{3}+f_{7}$ \\
$F18\textsubscript{Kepler}$ & 1.652501	& 0.00095	& 0.9946 & $f_{1}-f_{4}+f_{5}$ \\
\hline
\end{tabular}
\end{adjustbox}
\end{center}  
\end{table}

\begin{figure}
\begin{center}
\includegraphics[width=0.47\textwidth]{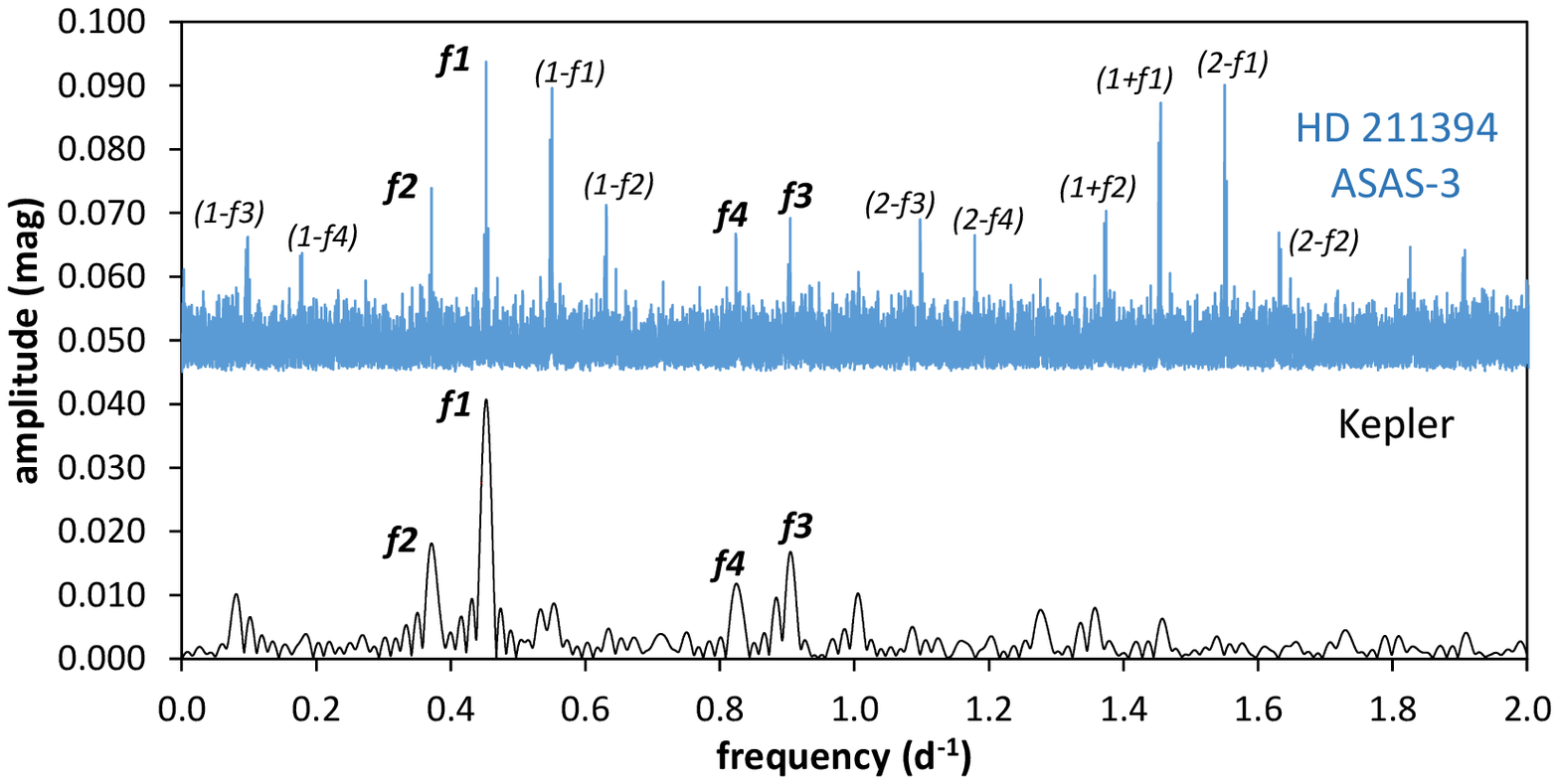}
\includegraphics[width=0.47\textwidth]{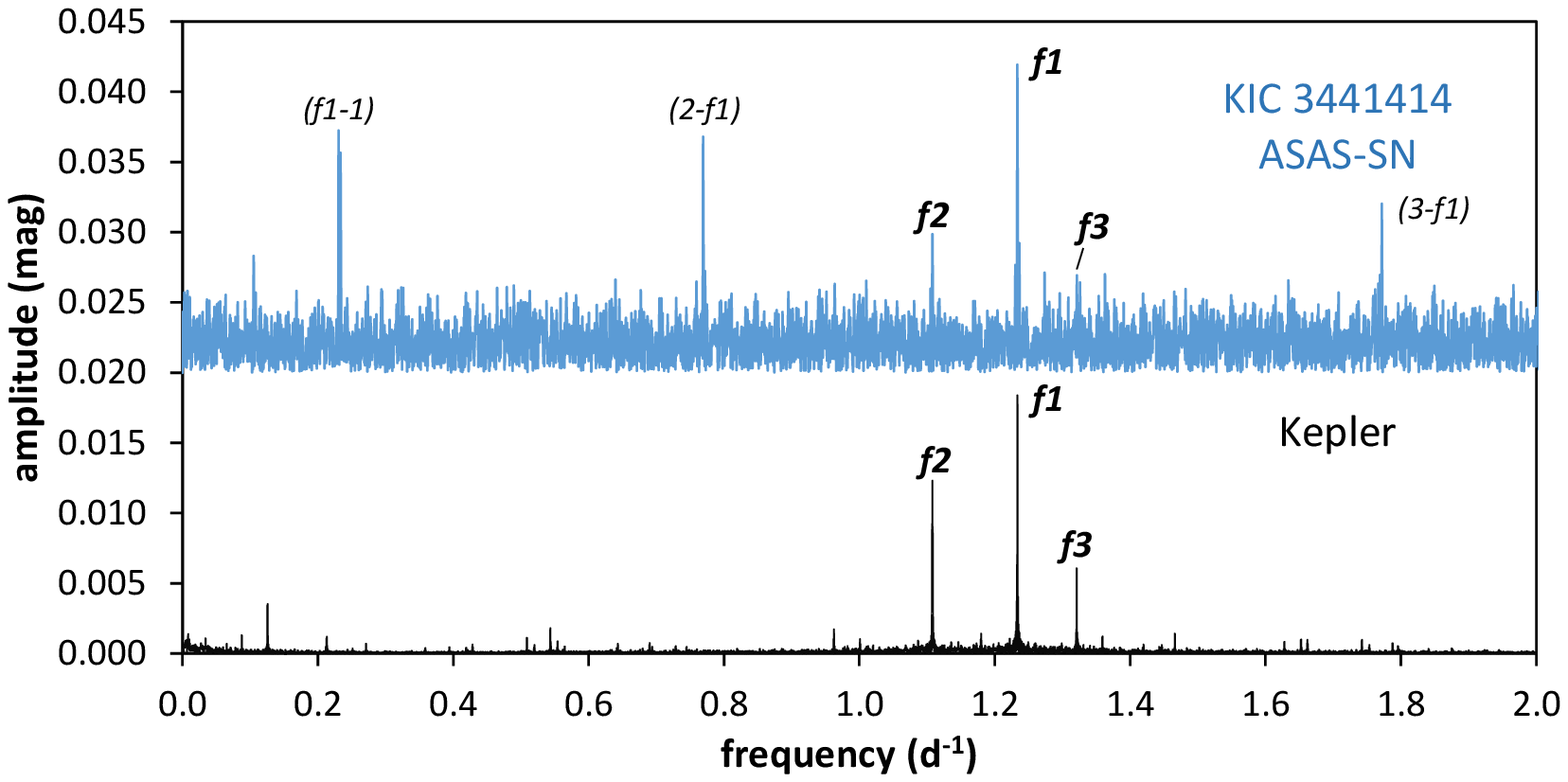}
\caption{Fourier amplitude spectra of HD\,211394, based on unwhitened ASAS-3 and $Kepler$ data (upper panel), and KIC\,3441414, based on unwhitened ASAS-SN and $Kepler$ data (lower panel). Note the presence of numerous alias peaks (identified using brackets) in the ground-based data.}
\label{fs_exemp}
\end{center}
\end{figure}

\subsection{Light variability pattern of the other HAGDOR variables} \label{discussion_other_HAGDORs}

Eight stars of our sample boast satellite photometry from $Kepler$ or TESS that allow a similarly detailed analysis. For the remaining seven objects with only ground-based observations, however, the precision and number of available measurements is not sufficient to perform an analysis of a vast number of frequencies. Nevertheless, it becomes obvious that all of our sample HAGDOR stars exhibit multiperiodic variability in a similar way to KIC\,8113425: in all objects, the beating of closely spaced frequencies results in total amplitudes that considerably exceed the amplitudes of the base frequencies. This leads to the observed 'upward trends' in the light curves, as described in \citet{kurtz15} and clearly seen for example in the light curves of KIC\,8113425 (Fig. \ref{fig1}, upper panel), HD\,211394 and KIC\,3441414 (both Fig. \ref{fig_light_curves1}). These upward trends can for example arise if the phase difference of the second harmonic in relation to the base frequency is nearly zero, as has been observed for $f_{4}$ in KIC\,8113425. Similar upward trends are clearly present in the light curves of all other investigated HAGDOR variables, although in ASAS-3 and ASAS-SN data, these are sometimes only represented by a 'smattering' of bright data points around the time of maximum light in the phase diagrams. All HAGDOR stars, therefore, belong to the ASYM group of \citet{balona11} (cf. also Section \ref{comparison}).

As examples, Fig. \ref{fig_light_curves1} illustrates the $Kepler$ light curves of the two HAGDOR stars HD\,211394 and KIC\,3441414. The large peak-to-peak amplitudes of the observed variations (0.324 and 0.122\,mag, respectively) become directly obvious. HD\,211394, in particular, is noteworthy because it is one of the stars with the largest amplitude in our sample. Using Amp($V$)\,/\,Amp($Kp$)\,=\,1.07(3) (cf. Sect. \ref{target_selection}), its light variations reach a peak-to-peak amplitude in $V$ of about 0.35\,mag, which is only rivalled by the variability of HD\,33575 (Amp($V$)\,=\,0.35\,mag).

The corresponding Fourier amplitude spectra are shown in Fig. \ref{fs_exemp}, the frequency solutions, as derived from the different data sources, are presented in Tables \ref{fs_example1} and \ref{fs_example2}. Fig. \ref{fs_exemp} also illustrates the higher noise level, lower sampling rate and the presence of one-day alias peaks in the ground-based ASAS-3 and ASAS-SN data. Nevertheless, these data span a much longer time baseline and are still very much suitable for the goals of the present investigation, which is also demonstrated by the good agreement between the principal frequencies derived from the different data sources. We also note that pulsation amplitudes are higher in $V$ than in the broad-band $Kepler$ and TESS data, which can also be seen in Fig. \ref{fs_exemp} (cf. Section \ref{target_selection}).

The frequency solutions for all stars are shown in the Appendix in Section \ref{frequency_solutions}. Fourier amplitude spectra of all HAGDOR stars are presented in Section \ref{fourier_spectra}.

\begin{figure*}
\begin{center}
 \mbox{
		\subfigure{\includegraphics[width=0.33\textwidth]{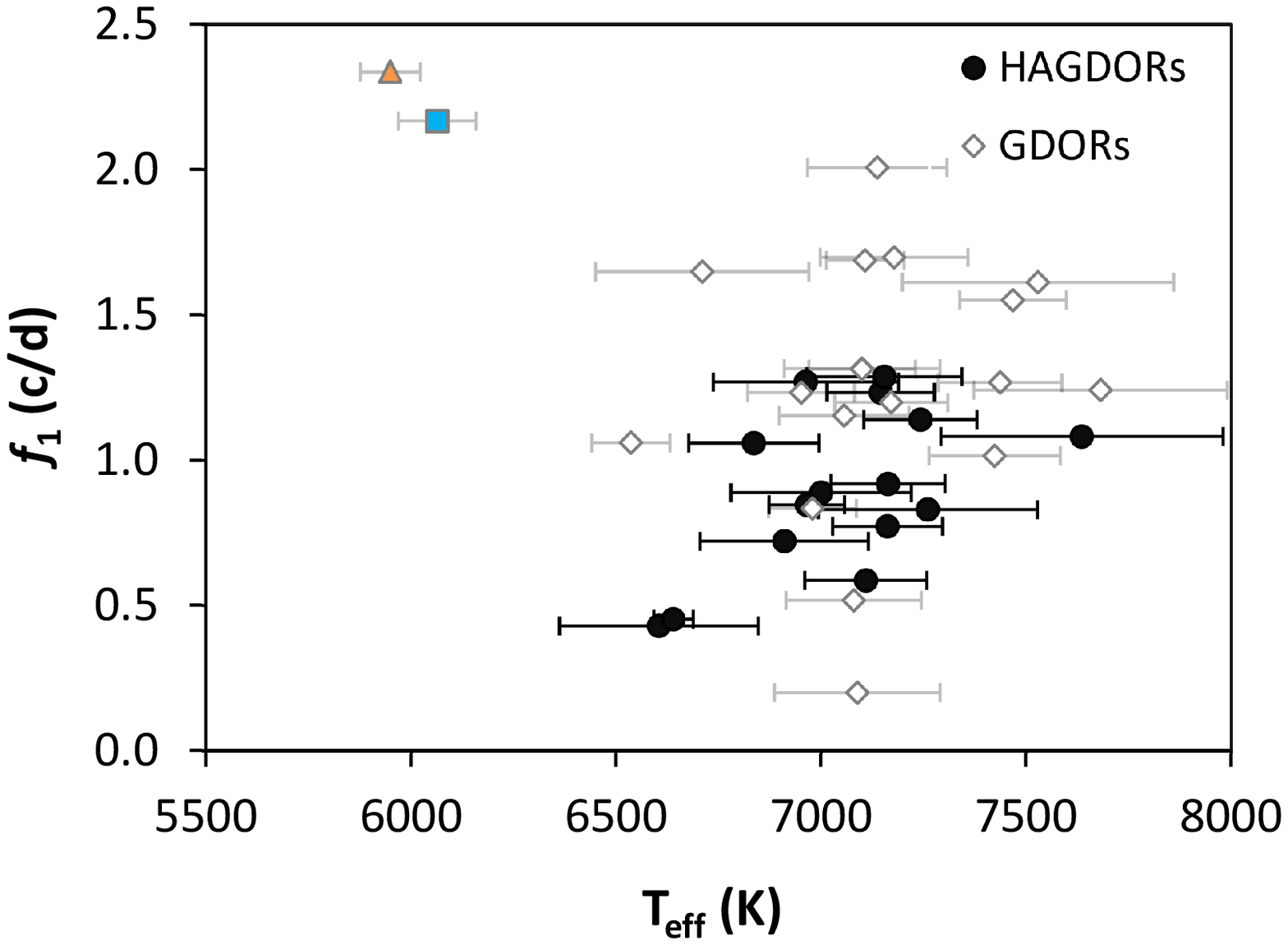}}
		\subfigure{\includegraphics[width=0.33\textwidth]{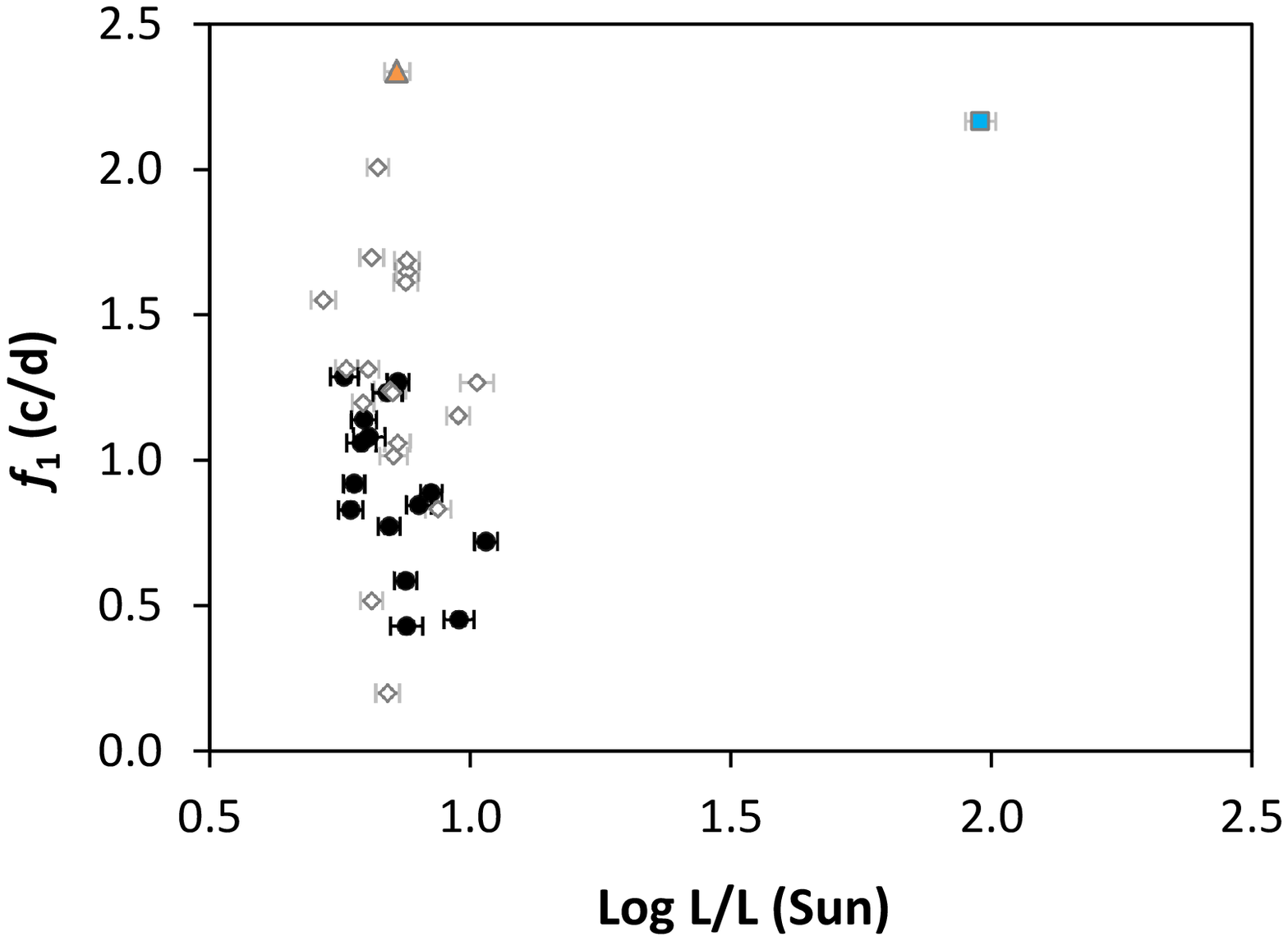}}
		\subfigure{\includegraphics[width=0.33\textwidth]{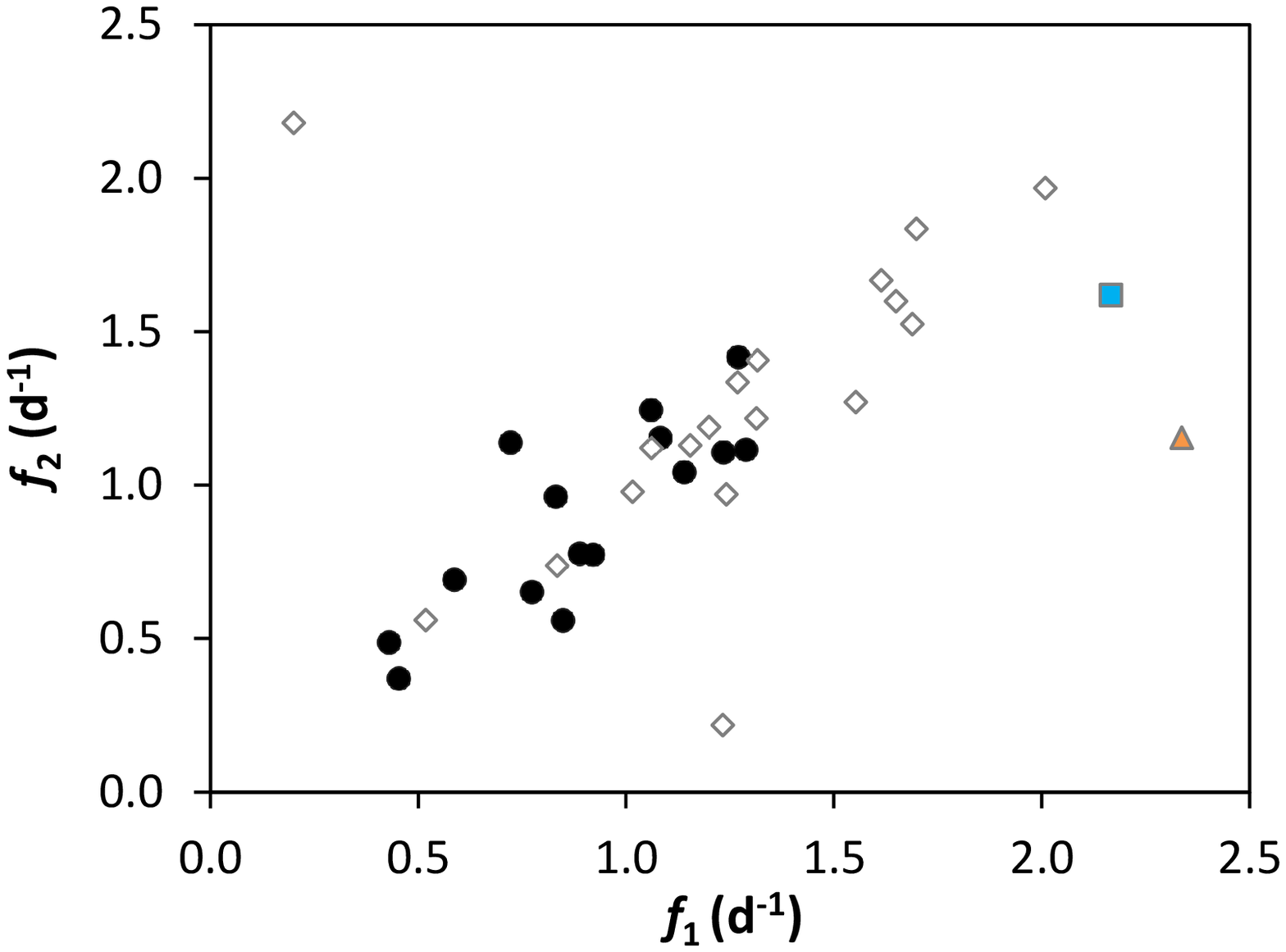}}
		}
 \mbox{
		\subfigure{\includegraphics[width=0.33\textwidth]{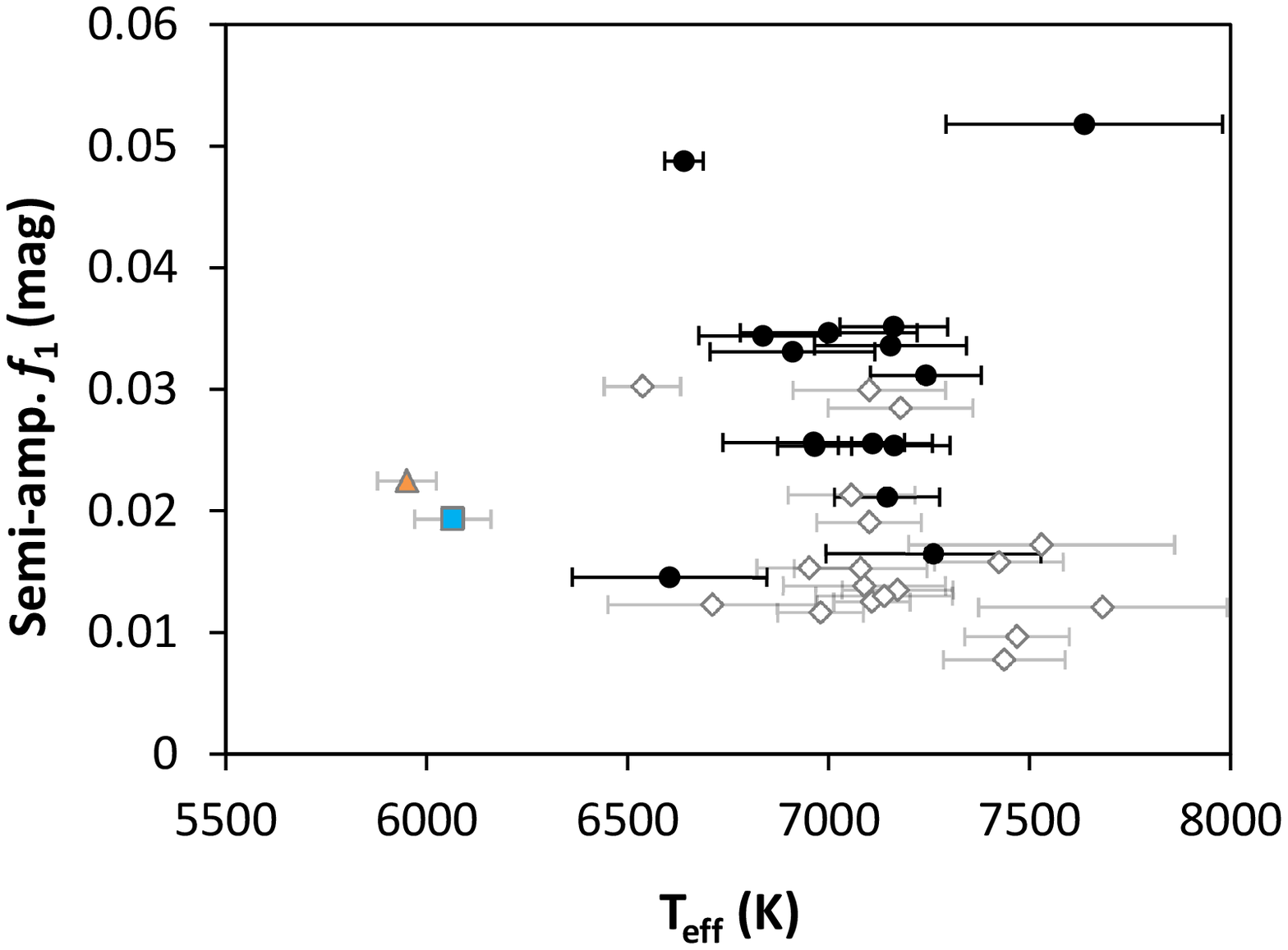}}
		\subfigure{\includegraphics[width=0.33\textwidth]{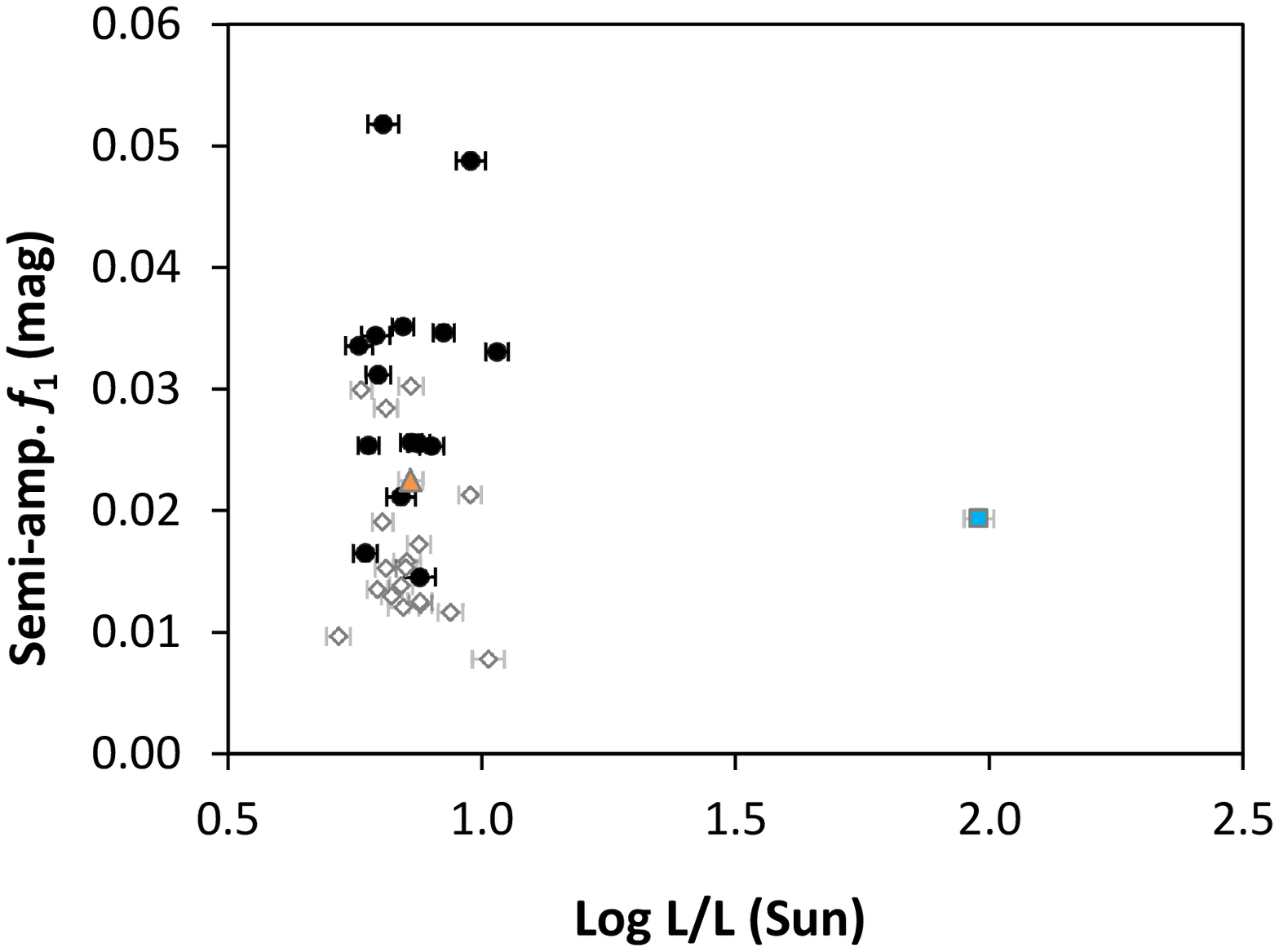}}
		\subfigure{\includegraphics[width=0.33\textwidth]{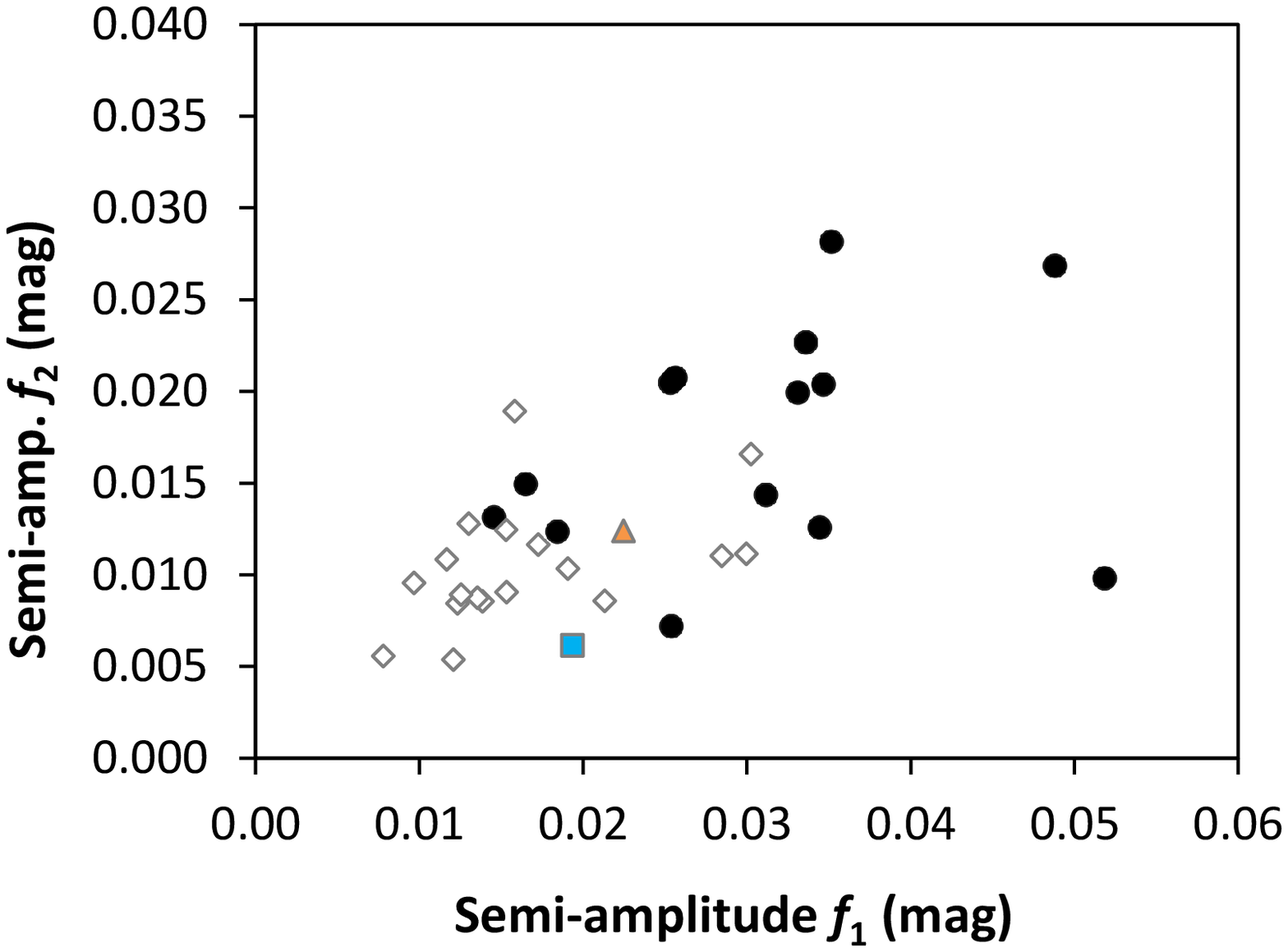}}
		}
\caption{Correlations between several observables for the sample of HAGDOR stars (black circles) and GDOR stars (open diamonds). The GDOR stars GSC 09289-02186 and GSC 04281-00186 are indicated, respectively, by an orange triangle and a blue rectangle in all plots. The panels investigate primary frequency versus effective temperature (upper left), semi-amplitude of the primary frequency versus effective temperature (lower left), primary frequency versus luminosity (upper middle), semi-amplitude of the primary frequency versus luminosity (lower middle), the correlation between the two most significant frequencies (upper right) and their corresponding semi-amplitudes (lower right). There generally is no significant difference in distribution between HAGDOR and GDOR stars, which do not seem to be physically distinct in any other respect than their total variability amplitudes.}
\label{correlations}
\end{center}
\end{figure*}

In summary, we conclude that the observed high amplitudes in the HAGDOR stars are caused by the presence of multiple, closely-spaced frequencies and their interactions. In this way, although the maximum amplitude of the primary frequencies does not exceed an amplitude of 0.1\,mag, total light variability amplitudes of over 0.3\,mag ($V$) can be attained, as for example in the case of HD\,211394 and HD\,33575. This is an interesting result that shows the need for the revision of the customary GDOR star class definition and provides important input for pulsational modeling attempts.

\subsection{Comparison of the samples of GDOR and HAGDOR variables} \label{comparison}

Several studies have indicated that high-amplitude DSCT (HADS) stars are distinguished from the lower-amplitude DSCT variables by several criteria: they generally show only one or two excited radial modes (usually the fundamental mode and/or first harmonic) and are mostly slow rotators, which seems to be a requirement for the observed high-amplitude pulsation \citep{breger00,mcnamara00}. It has also been postulated that HADS stars are in an evolutionary stage that puts them between low-mass classical Cepheids and high-mass DSCT stars; however, in their investigation of the physical nature of HADS stars using $Kepler$ data, \citet{balona16} rejected this scenario and found that HADS stars are distributed randomly across the DSCT instability strip. No physical attribute was found that separates HADS stars from their low-amplitude counterparts, although there seems to be a general tendency for the number of combination frequencies to increase with increasing amplitude of the parent frequencies \citep{balona16}. Further investigation into the relationship between DSCT, HADS stars and the related low-metallicity SX Phe stars is clearly desirable.

Here we investigate the relationship between HAGDOR and GDOR stars. Their locations in the instability strip (Fig. \ref{fig_instability_strip}) and \logTeff\ versus \logl\ diagram (cf. Section \ref{astrophysical_parameters}) suggest that GDOR and HAGDOR stars are not physically distinct objects but rather a homogeneous group of variables. To further tackle this question, we have investigated the relation between the primary variability frequency and the parameters effective temperature and luminosity (Fig. \ref{correlations}). Both GDOR and HAGDOR stars overlap in the investigated parameter spaces and no significant correlation was found. The situation is similar for the semi-amplitudes of the primary frequencies: although HAGDOR stars tend to show larger amplitudes, as expected, no distinct boundary between GDOR and HAGDOR stars is observed but rather a gradual transition and considerable overlap. We have also correlated the two most significant frequencies and their corresponding semi-amplitudes (Fig. \ref{correlations}, right panels). Again, while HAGDOR stars tend to show larger amplitudes, no significant differences were found between the two investigated groups. Interestingly, nearly all of the investigated stars show closely-spaced primary frequencies, which seems to be a characteristic of the class of GDOR variables.

It is interesting to point out, though, that all HAGDOR stars have light curves of the ASYM type, whereas the control sample of GDORs is predominately made up of stars having SYM light curves. The occurrence of the characteristic beating effects observed in the ASYM group, therefore, seems to be necessary for the development of the high variability amplitudes observed in the HAGDOR stars.

In summary, the here presented evidence -- although based on rather small sample sizes -- suggests that GDORs and HAGDORs are not physically distinct in any other respect than their total variability amplitudes but merely represent the low- and high-amplitude ends of the same, uniform group of variables, with a continuous progression from low to high total amplitudes.

The G0\,V star GSC\,09289-02186 and the G-type giant GSC\,04281-00186 (cf. Section \ref{astrophysical_parameters}), both belonging to the GDOR star sample, constitute the most obvious outliers in the diagrams presented in Figure \ref{correlations}, in particular in the primary frequency versus effective temperature plot (upper left panel), and deserve special mention. These stars exhibit the characteristic variability patterns of GDOR stars and show light curves of the SYM type; however, both stars are situated outside the traditional boundaries of the GDOR realm, whose red border is found at a spectral type of approximately F7 (cf. Section \ref{introduction}). Furthermore, GDOR stars belong to luminosity classes IV or V by definition \citep{kaye99}. The case of a giant star exhibiting pulsation compatible with a GDOR type is therefore of interest.

\begin{figure}
\begin{center}
\includegraphics[width=0.47\textwidth]{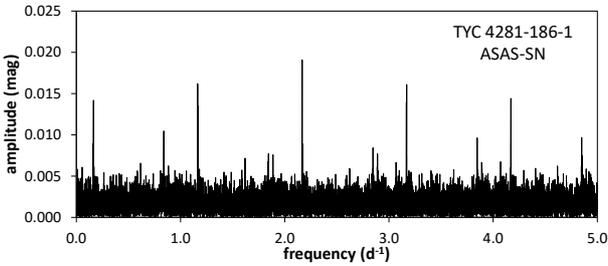}
\caption{Fourier amplitude spectrum of the G-type giant pulsator GSC\,04281-00186, based on unwhitened ASAS-SN data. The y-axis denotes semi-amplitudes, as derived with PERIOD04.}
\label{fig_fs_gsc4281_186}
\end{center}
\end{figure}

\citet{balona11} noticed the occurrence of giant stars with $T_{\mathrm{eff}}$\,$<$\,6000\,K in their sample of GDOR stars with MULT-type light curves. They surmised that these objects represent solar-type oscillators among late G giants. However, the variability pattern of the giant star GSC\,04281-00186, which is characterized by two main pulsations frequencies with rather high amplitudes (Fig. \ref{fig_fs_gsc4281_186}), is vastly different from the variability seen in solar-like pulsating stars, which show rich frequency spectra of stochastically excited modes (e.g. \citealt{bedding10,chaplin13}).

Interestingly, GSC\,09289-02186 and GSC\,04281-00186 exhibit the highest principal pulsation frequencies of all our sample stars (2.33692\,d$^{-1}$ and 2.16673\,d$^{-1}$, respectively). Further detailed studies of these objects are encouraged, which might shed light on their pulsational properties and the mechanisms at work and could, perhaps, lead to another expansion of the class defintion of GDOR stars. This, however, is out of the scope of the present investigation.

%%%%%%%%%%%%%%%%%%%%%%%%%%%%%%%%%%%%%%%%%%%%%%%%%%%%%%%%%%%%%

\section{Conclusion} \label{conclusion}

We have investigated high-amplitude GDOR (HAGDOR) stars showing light variability amplitudes well beyond the traditional 0.1\,mag ($V$) limit, with the aim of unraveling the mechanisms behind the observed high amplitudes and investigating whether these objects are in any way physically distinct from regular GDOR stars ($V$\textsubscript{amp}$\,\le\,$0.1\,mag). To this end, a sample of 15 HAGDOR stars and a control sample of 20 regular GDOR stars boasting extensive photometric time series data were collected. 

As a first step, we calculated astrophysical parameters and investigated our sample stars in the \logTeff\ versus \logl\ diagram. No significant differences between the location of the HAGDOR stars and the GDOR stars were found -- both groups are well distributed over the whole main sequence up to \logl\,$<$\,1.1. Employing publicly available survey data ($Kepler$, ASAS-3, ASAS-SN and, in the case of one star, TESS data) and our own observations, we analyzed the photometric variability of our target stars, using the well-described HAGDOR star KIC\,8113425 \citep{kurtz15} as a model case.

We found that all HAGDOR variables show light curves of the ASYM type and behave similarly to KIC\,8113425 in that they exhibit multiple frequencies whose beating results in total amplitudes considerably exceeding the amplitudes of the base frequencies. Hence, the high amplitudes observed in the HAGDOR stars can be explained by the superposition of several base frequencies in interaction with their combination and overtone frequencies. Although the maximum amplitude of the primary frequencies does not exceed an amplitude of 0.1\,mag, total light variability amplitudes of over 0.3\,mag ($V$) can be attained in this way -- important input for pulsational modeling attempts.  We conclude that a revision of the traditional amplitude cut-off of 0.1\,mag ($V$) for GDOR stars is necessary. Based on the analyses of the present investigation, we propose a new cut-off value of 0.35\,mag ($V$). We caution, however, that this conclusion has been based on the analysis of a small sample and HAGDOR stars with larger amplitudes may be found subsequently.

To tackle the question whether HAGDOR stars and GDOR stars are physically distinct objects, correlations between the observed primary frequencies, amplitudes and other parameters like effective temperature and luminosity were investigated. HAGDOR stars tend to show larger amplitudes, and nearly all of the investigated stars exhibit closely-spaced primary frequencies, which seems to be a general characteristic of the class of GDOR variables. Apart from that, however, no significant differences were found; instead, both groups overlap and show gradual transitions in the investigated parameter spaces. We therefore conclude that low- and high-amplitude GDOR stars are not physically distinct in any other respect than their total variability amplitudes but merely represent two ends of the same, uniform group of variables. However, we caution that the sample sizes used for our investigation are small (15 HAGDOR stars and a comparison sample of 20 regular GDOR stars with total variability amplitudes of 0.05\,$\le$$V$$\le\,$0.1\,mag); our conclusions, therefore, should be confirmed using a larger sample of stars with good photometric time series observations.

We call attention to the GDOR variables GSC\,09289-02186 and GSC\,04281-00186, which exhibit the highest principal pulsation frequencies of our sample stars. Furthermore, both stars are situated outside the traditional boundaries of the GDOR realm and constitute the most obvious outliers in the investigated parameter spaces. Although some discrepancies in the available data exist, according to its calibrated astrophysical parameters, GSC\,04281-00186 is a G-type giant, which is intriguing as GDOR stars belong to luminosity classes IV or V by definition. Further detailed studies of these objects are encouraged.

%%%%%%%%%%%%%%%%%%%%%%%%%%%%%%%%%%%%%%%%%%%%%%%%%%%%%%%%%%%%

\section*{Acknowledgements}
We thank the referee for his/her thoughtful report that helped to improve the paper. This research has made use of the SIMBAD and VizieR databases operated at the Centre de Donn\'{e}es Astronomiques (Strasbourg) in France, of the AAVSO International Variable Star Index (VSX) and of the Two Micron All Sky Survey (2MASS), which is a joint project of the University of Massachusetts and the Infrared Processing and Analysis Center/California Institute of Technology, funded by the National Aeronautics and Space Administration and the National Science Foundation. This paper includes data collected by the $Kepler$ mission. Funding for the $Kepler$ mission is provided by the NASA Science Mission directorate. This paper includes data collected with the TESS mission, obtained from the MAST data archive at the Space Telescope Science Institute (STScI). Funding for the TESS mission is provided by the NASA Explorer Program. STScI is operated by the Association of Universities for Research in Astronomy, Inc., under NASA contract NAS 5–26555. This work has also made use of data from the European Space Agency (ESA) mission Gaia (http://www.cosmos.esa.int/gaia), processed by the Gaia Data Processing and Analysis Consortium (DPAC, http://www.cosmos.esa.int/web/gaia/dpac/consortium). Funding for the DPAC has been provided by national institutions, in particular the institutions participating in the Gaia Multilateral Agreement. Guoshoujing Telescope (the Large Sky Area Multi-Object Fiber Spectroscopic Telescope LAMOST) is a National Major Scientific Project built by the Chinese Academy of Sciences. Funding for the project has been provided by the National Development and Reform Commission. LAMOST is operated and managed by the National Astronomical Observatories, Chinese Academy of Sciences.

%%%%%%%%%%%%%%%%%%%%%%%%%%%%%%%%%%%%%%%%%%%%%%%%%%%%%%%%%%%%

\section*{Data Availability}
The data underlying this article will be shared on reasonable request to the corresponding author.


\bibliographystyle{mnras}
\bibliography{HAGDORS}

%%%%%%%%%%%%%%%%%%%%%%%%%%%%%%%%%%%%%%%%%%%%%%%%%%
%%%%%%%%%%%%%%%%%%%% APPENDIX %%%%%%%%%%%%%%%%%%
%%%%%%%%%%%%%%%%%%%%%%%%%%%%%%%%%%%%%%%%%%%%%%%%%%

\appendix

%%%%%%%%%%%%%%%%%%%%%%%%%%%%%%%%%%%%%%%%%%%%%%%%%%%%%%%%%%%%%

\section{Frequency solutions} \label{frequency_solutions}

In the following, the frequency solutions for all sample stars are presented in tabular form. Stars are listed in order of increasing right ascension. In each table, the columns denote: (1) frequency number; (2) frequency value; (3) semi-amplitude; (4) corresponding phase; (5) frequency identification. All values have been derived using \textsc{PERIOD04}, as outlined in Section \ref{data_analysis}.

\clearpage

\begin{table}
\caption{Frequency solution for \textbf{GSC\,02831-00348} (V0758\,And).}
\label{fs1}
\begin{center}
\begin{tabular}{lllll}
\hline
\hline
(1) & (2) & (3) & (4) & (5) \\
Freq.No. & Frequency & Amplitude & Phase & ID \\
\hline
$F1\textsubscript{ASAS-SN}$ & 1.081484 & 0.052 & 0.6574 & $f_{1}$ \\
$F2\textsubscript{ASAS-SN}$ & 1.155404 & 0.010 & 0.7579 & $f_{2}$ \\
\hline
\end{tabular}                                                                                                                                                             
\end{center}  
\end{table}

\begin{table}
\caption{Frequency solution for \textbf{HD\,17721} (HIP\,13089).}
\label{fs2}
\begin{center}
\begin{tabular}{lllll}
\hline
\hline
(1) & (2) & (3) & (4) & (5) \\
Freq.No. & Frequency & Amplitude & Phase & ID \\
\hline
$F1\textsubscript{ASAS-3}$ & 0.919238	& 0.025	& 0.3389 & $f_{1}$ \\
$F2\textsubscript{ASAS-3}$ & 0.774241	& 0.007	& 0.2701 & $f_{2}$ \\
\hline
$F1\textsubscript{TESS}$ & 0.919038	& 0.0192	& 0.2682 & $f_{1}$ \\
$F2\textsubscript{TESS}$ & 0.776222	& 0.0090	& 0.9202 & $f_{2}$ \\								
$F3\textsubscript{TESS}$ & 0.653821	& 0.0057	& 0.8019 & $f_{3}$ \\		
$F4\textsubscript{TESS}$ & 1.120093	& 0.0056	& 0.5962 & $f_{4}$ \\		
$F5\textsubscript{TESS}$ & 0.196703	& 0.0045	& 0.5247 & $f_{5}$ \\
\hline
\end{tabular}                                                                                                                                                             
\end{center}  
\end{table}

\begin{table}
\caption{Frequency solution for \textbf{HD\,33575} (NSV\,1858).}
\label{fs3}
\begin{center}
\begin{tabular}{lllll}
\hline
\hline
(1) & (2) & (3) & (4) & (5) \\
Freq.No. & Frequency & Amplitude & Phase & ID \\
\hline
$F1\textsubscript{ASAS-3}$ & 0.773239	& 0.032	& 0.3632 & $f_{1}$ \\
$F2\textsubscript{ASAS-3}$ & 0.652773	& 0.025	& 0.9155 & $f_{2}$ \\
$F3\textsubscript{ASAS-3}$ & 1.425986	& 0.024	& 0.6162 & $f_{1}+f_{2}$ \\
\hline
$F1\textsubscript{ASAS-SN}$ & 0.773220	& 0.044	& 0.6462 & $f_{1}$ \\
$F2\textsubscript{ASAS-SN}$ & 0.652721	& 0.033	& 0.1973 & $f_{2}$ \\
\hline
\end{tabular}                                                                                                                                                             
\end{center}  
\end{table}

\begin{table}
\caption{Frequency solution for \textbf{HD\,50875} (NSV\,3272).}
\label{fs4}
\begin{center}
\begin{tabular}{lllll}
\hline
\hline
(1) & (2) & (3) & (4) & (5) \\
Freq.No. & Frequency & Amplitude & Phase & ID \\
\hline
$F1\textsubscript{ASAS-3}$ & 0.586397 & 0.026	& 0.2689 & $f_{1}$ \\
$F2\textsubscript{ASAS-3}$ & 0.692966 & 0.021	& 0.5241 & $f_{2}$ \\
$F3\textsubscript{ASAS-3}$ & 0.110471 & 0.011	& 0.9279 & $f_{2}-f_{1}$? \\
\hline
\end{tabular}                                                                                                                                                             
\end{center}  
\end{table}

\begin{table}
\caption{Frequency solution for \textbf{HD\,85693} (NSV\,18291).}
\label{fs5}
\begin{center}
\begin{tabular}{lllll}
\hline
\hline
(1) & (2) & (3) & (4) & (5) \\
Freq.No. & Frequency & Amplitude & Phase & ID \\
\hline
$F1\textsubscript{ASAS-3}$ & 0.888140 & 0.034	& 0.8495 & $f_{1}$ \\
$F2\textsubscript{ASAS-3}$ & 0.778475	& 0.021	& 0.9916 & $f_{2}$ \\
$F3\textsubscript{ASAS-3}$ & 1.666578	& 0.014	& 0.2639 & $f_{1}+f{2}$\\
$F4\textsubscript{ASAS-3}$ & 0.793137	& 0.012	& 0.0540 & $f_{3}$ \\
\hline
\end{tabular}                                                                                                                                                             
\end{center}  
\end{table}

\begin{table}
\caption{Frequency solution for \textbf{GSC\,09046-00646} (ASAS\,J163451-6446.3).}
\label{fs6}
\begin{center}
\begin{tabular}{lllll}
\hline
\hline
(1) & (2) & (3) & (4) & (5) \\
Freq.No. & Frequency & Amplitude & Phase & ID \\
\hline
$F1\textsubscript{ASAS-3}$ & 0.846993	& 0.025	& 0.9727 & $f_{1}$ \\
$F2\textsubscript{ASAS-3}$ & 0.560801	& 0.020	& 0.1025 & $f_{2}$ \\
\hline
$F1\textsubscript{ASAS-SN}$ & 0.560868	& 0.023	& 0.3169 & $f_{2}$ \\
$F2\textsubscript{ASAS-SN}$ & 0.847028	& 0.024	& 0.0205 & $f_{1}$ \\
\hline
\end{tabular}                                                                                                                                                             
\end{center}  
\end{table}

\begin{table}
\caption{Frequency solution for \textbf{HD\,150538} (NSV\,20738).}
\label{fs7}
\begin{center}
\begin{tabular}{lllll}
\hline
\hline
(1) & (2) & (3) & (4) & (5) \\
Freq.No. & Frequency & Amplitude & Phase & ID \\
\hline
$F1\textsubscript{ASAS-3}$ & 0.721031	& 0.033	& 0.4240 & $f_{1}$ \\
$F2\textsubscript{ASAS-3}$ & 1.140350	& 0.020	& 0.9765 & $f_{2}$ \\
$F3\textsubscript{ASAS-3}$ & 1.044124	& 0.013	& 0.0028 & $f_{3}$ \\
\hline
\end{tabular}                                                                                                                                                             
\end{center}  
\end{table}

\begin{table}
\caption{Frequency solution for \textbf{KIC\,3847822} (TYC\,3134-2121-1).}
\label{fs8}
\begin{center}
\begin{tabular}{lllll}
\hline
\hline
(1) & (2) & (3) & (4) & (5) \\
Freq.No. & Frequency & Amplitude & Phase & ID \\
\hline
$F1\textsubscript{ASAS-SN}$ & 0.829894	& 0.018	& 0.5224 & $f_{1}$ \\
$F2\textsubscript{ASAS-SN}$ & 0.963088	& 0.017 & 0.3484 & $f_{2}$ \\
$F3\textsubscript{ASAS-SN}$ & 0.915570	& 0.011 & 0.0795 & $f_{2}$ \\
\hline
$F1\textsubscript{Kepler}$ & 0.829875	& 0.01600	& 0.6923 & $f_{1}$ \\
$F2\textsubscript{Kepler}$ & 0.963154 & 0.01494 &	0.7443 & $f_{2}$ \\
$F3\textsubscript{Kepler}$ & 0.915471 &	0.00771 &	0.8871 & $f_{3}$ \\
$F4\textsubscript{Kepler}$ & 1.096999 &	0.00762 &	0.9753 & $f_{4}$ \\
$F5\textsubscript{Kepler}$ & 0.133330	& 0.00673 &	0.5757 & $f_{2}-f_{1}$ \\
$F6\textsubscript{Kepler}$ & 0.085594 &	0.00347 &	0.0515 & $f_{3}-f_{1}$ \\
$F7\textsubscript{Kepler}$ & 0.047627 & 0.00345 & 0.1426 & $f_{2}-f_{3}$ \\
$F8\textsubscript{Kepler}$ & 0.267117 &	0.00339 &	0.1179 & $f_{4}-f_{1}$ \\
$F9\textsubscript{Kepler}$ & 0.133790 &	0.00420 &	0.3820 & $f_{4}-f_{2}$ \\
$F10\textsubscript{Kepler}$ & 0.781546 & 0.00277 & 0.0652 & $f_{5}$ \\
$F11\textsubscript{Kepler}$ & 0.181559 & 0.00259 & 0.7470 & $f_{4}-f_{3}$ \\
\hline
\end{tabular}                                                                                                                                                             
\end{center}  
\end{table}

\begin{table}
\caption{Frequency solution for \textbf{KIC\,3441414} (GSC\,03134-00901).}
\label{fs9}
\begin{center}
\begin{adjustbox}{max width=0.5\textwidth}
\begin{tabular}{lllll}
\hline
\hline
(1) & (2) & (3) & (4) & (5) \\
Freq.No. & Frequency & Amplitude & Phase & ID \\
\hline
$F1\textsubscript{ASAS-SN}$ & 1.233502	& 0.021	& 0.5729 & $f_{1}$ \\
$F2\textsubscript{ASAS-SN}$ & 1.107843	& 0.010	& 0.7617 & $f_{2}$ \\
$F3\textsubscript{ASAS-SN}$ & 1.321001	& 0.007	& 0.2928 & $f_{3}$ \\
\hline
$F1\textsubscript{Kepler}$ & 1.233479	& 0.01841	& 0.1633 & $f_{1}$ \\	
$F2\textsubscript{Kepler}$ & 1.107875	& 0.01237	& 0.8904 & $f_{2}$ \\
$F3\textsubscript{Kepler}$ & 1.320999	& 0.00622	& 0.6805 & $f_{3}$ \\
$F4\textsubscript{Kepler}$ & 0.125600 & 0.00350	& 0.1636 & $f_{1}-f_{2}$ \\
$F5\textsubscript{Kepler}$ & 2.341354	& 0.00230	& 0.3444 & $f_{1}+f_{2}$ \\
$F6\textsubscript{Kepler}$ & 0.543345	& 0.00176	& 0.3657 & $f_{4}$ \\	
$F7\textsubscript{Kepler}$ & 0.962388	& 0.00170	& 0.4148 & $f_{5}$ \\
$F8\textsubscript{Kepler}$ & 2.466959	& 0.00164 & 0.6123 & $2f_{1}$	\\
$F9\textsubscript{Kepler}$ & 1.465986	& 0.00138	& 0.7528 & $f_{6}$ \\
$F10\textsubscript{Kepler}$ & 0.087513	& 0.00134	& 0.4919 & $f_{3}-f_{1}$ \\
$F11\textsubscript{Kepler}$ & 1.179790	& 0.00130	& 0.4088 & $f_{7}$ \\	
$F12\textsubscript{Kepler}$ & 0.213140	& 0.00120	& 0.6362 & $f_{3}-f_{2}$ \\
$F13\textsubscript{Kepler}$ & 0.508685	& 0.00109	& 0.6989 & $f_{8}$ \\
$F14\textsubscript{Kepler}$ & 1.661927	& 0.00109	& 0.3922 & $f_{9}$ \\
$F15\textsubscript{Kepler}$ & 1.742302	& 0.00108	& 0.0020 & $f_{10}$ \\
$F16\textsubscript{Kepler}$ & 1.359081	& 0.00105	& 0.6474 & $2f_{1}-f_{2}$ \\
$F17\textsubscript{Kepler}$ & 0.034023	& 0.00102	& 0.4082 & $-2f_{1}+f_{3}+f_{7}$ \\
$F18\textsubscript{Kepler}$ & 1.652501	& 0.00095	& 0.9946 & $f_{1}-f_{4}+f_{5}$ \\
\hline
\end{tabular}
\end{adjustbox}
\end{center}  
\end{table}
						
\begin{table}
\caption{Frequency solution for \textbf{KIC\,7448050} (ASAS\,J193103+4302.1).}
\label{fs10}
\begin{center}
\begin{adjustbox}{max width=0.5\textwidth}
\begin{tabular}{lllll}
\hline
\hline
(1) & (2) & (3) & (4) & (5) \\
Freq.No. & Frequency & Amplitude & Phase & ID \\
\hline
$F1\textsubscript{ASAS-SN}$ & 1.139451 & 0.031 & 0.9378 & $f_{1}$ \\
$F2\textsubscript{ASAS-SN}$ & 1.043578 & 0.014 & 0.3619 & $f_{2}$ \\
$F3\textsubscript{ASAS-SN}$ & 1.560555 & 0.013 & 0.7565 & $f_{3}$ \\
$F4\textsubscript{ASAS-SN}$ & 2.278806 & 0.006 & 0.8760 & $2f_{1}$ \\
\hline
$F1\textsubscript{Kepler}$ & 1.139452 & 0.02766 & 0.9283 & $f_{1}$ \\
$F2\textsubscript{Kepler}$ & 1.043526 & 0.01312 & 0.7471 & $f_{2}$ \\
$F3\textsubscript{Kepler}$ & 1.560513 & 0.00949 & 0.0753 & $f_{3}$ \\
$F4\textsubscript{Kepler}$ & 1.282439 & 0.00790 & 0.7678 & $f_{4}$ \\
$F5\textsubscript{Kepler}$ & 0.421057 & 0.00404 & 0.0003 & $f_{3}-f_{1}$ \\
$F6\textsubscript{Kepler}$ & 1.785288 & 0.00347 & 0.2634 & $f_{5}$ \\
$F7\textsubscript{Kepler}$ & 0.237194 & 0.00311 & 0.8573 & $f_{4}-f_{2}$ \\
$F8\textsubscript{Kepler}$ & 2.182984 & 0.00298 & 0.9297 & $f_{1}+f_{2}$ \\
$F9\textsubscript{Kepler}$ & 1.376648 & 0.00297 & 0.0357 & $f_{6}$ \\
$F10\textsubscript{Kepler}$ & 1.399469 & 0.00279 & 0.0933 & $f_{7}$ \\
$F11\textsubscript{Kepler}$ & 0.142968 & 0.00277 & 0.8800 & $f_{4}-f_{1}$ \\
\hline
\end{tabular}
\end{adjustbox}
\end{center}  
\end{table}

\begin{table}
\caption{Frequency solution for \textbf{KIC\,6953103} (2MASSJ19325124+4228465).}
\label{fs11}
\begin{center}
\begin{adjustbox}{max width=0.5\textwidth}
\begin{tabular}{lllll}
\hline
\hline
(1) & (2) & (3) & (4) & (5) \\
Freq.No. & Frequency & Amplitude & Phase & ID \\
\hline
$F1\textsubscript{Kepler}$ & 1.287599 & 0.03357 & 0.7980 & $f_{1}$ \\
$F2\textsubscript{Kepler}$ & 1.115789 & 0.02269 & 0.5773 & $f_{2}$ \\
$F3\textsubscript{Kepler}$ & 1.198750 & 0.02199 & 0.1149 & $f_{3}$ \\
$F4\textsubscript{Kepler}$ & 0.171818 & 0.00959 & 0.0547 & $f_{1}-f_{2}$ \\
$F5\textsubscript{Kepler}$ & 0.088856 & 0.00802 & 0.6031 & $f_{1}-f_{3}$ \\
$F6\textsubscript{Kepler}$ & 1.026945 & 0.00601 & 0.7031 & $f_{2}+f_{3}-f_{1}$	\\
$F7\textsubscript{Kepler}$ & 2.403390 & 0.00534 & 0.6671 & $f_{1}+f_{2}$ \\
$F8\textsubscript{Kepler}$ & 1.254955 & 0.00517 & 0.4008 & $f_{4}$ \\
$F9\textsubscript{Kepler}$ & 0.071200 & 0.00505 & 0.3261 & $f_{5}-f_{1}$ \\
$F10\textsubscript{Kepler}$ & 1.358803 & 0.00461 & 0.3727 & $f_{5}$ \\
$F11\textsubscript{Kepler}$ & 2.486359 & 0.00467 & 0.1549 & $f_{1}+f_{3}$ \\
$F12\textsubscript{Kepler}$ & 1.376442 & 0.00452 & 0.7867 & $2f_{1}-f_{3}$ \\
$F13\textsubscript{Kepler}$ & 2.575187 & 0.00428 & 0.9824 & $2f_{1}$ \\
$F14\textsubscript{Kepler}$ & 2.314550 & 0.00386 & 0.8803 & $f_{2}+f_{3}$ \\
$F15\textsubscript{Kepler}$ & 1.186985 & 0.00374 & 0.1643 & $-f_{1}+f_{2}+f_{5}$ \\
$F16\textsubscript{Kepler}$ & 1.459425 & 0.00344 & 0.1386 & $2f_{1}-f_{2}$ \\
$F17\textsubscript{Kepler}$ & 0.082985 & 0.00338 & 0.2898 & $-f_{2}+f_{3}$ \\
$F18\textsubscript{Kepler}$ & 0.943990 & 0.00319 & 0.8789 & $-f_{1}+2f_{2}$ \\
$F19\textsubscript{Kepler}$ & 2.474589 & 0.00306 & 0.2586 & $f_{2}+f_{5}$ \\
$F20\textsubscript{Kepler}$ & 1.132902 & 0.00305 & 0.4786 & $f_{6}$ \\
\hline
\end{tabular}
\end{adjustbox}
\end{center}  
\end{table}

\begin{table}
\caption{Frequency solution for \textbf{KIC\,8113425} (2MASSJ19474808+4354257).}
\label{fs12}
\begin{center}
\begin{adjustbox}{max width=0.5\textwidth}
\begin{tabular}{lllll}
\hline
\hline
(1) & (2) & (3) & (4) & (5) \\
Freq.No. & Frequency & Amplitude & Phase & ID \\
\hline
$F1\textsubscript{ASAS-SN}$ & 0.429956 & 0.014 & 0.7265 & $f_{1}$ \\
$F2\textsubscript{ASAS-SN}$ & 0.489422 & 0.011 & 0.2704 & $f_{2}$ \\
$F3\textsubscript{ASAS-SN}$ & 0.449965 & 0.008 & 0.2736 & $f_{3}$ \\
\hline
$F1\textsubscript{Kepler}$ & 0.430055 & 0.01428 & 0.5268 & $f_{1}$ \\
$F2\textsubscript{Kepler}$ & 0.489411 & 0.01285 & 0.9223 & $f_{2}$ \\
$F3\textsubscript{Kepler}$ & 0.450101 & 0.00990 & 0.1920 & $f_{3}$ \\
$F4\textsubscript{Kepler}$ & 0.461260 & 0.00763 & 0.4039 & $f_{4}$ \\
$F5\textsubscript{Kepler}$ & 0.919451 & 0.00706 & 0.7664 & $f_{1}+f_{2}$ \\
$F6\textsubscript{Kepler}$ & 0.950674 & 0.00636 & 0.4393 & $f_{2}+f_{4}$ \\
$F7\textsubscript{Kepler}$ & 0.891334 & 0.00483 & 0.9490 & $f_{1}+f_{4}$ \\
$F8\textsubscript{Kepler}$ & 0.059311 & 0.00445 & 0.6557 & $f_{2}-f_{1}$ \\
$F9\textsubscript{Kepler}$ & 0.978781 & 0.00418 & 0.3443 & $2f_{2}$ \\
$F10\textsubscript{Kepler}$ & 0.008106 & 0.00352 & 0.8156 & $f_{1}+f_{2}-f_{3}-f_{4}$ \\
$F11\textsubscript{Kepler}$ & 0.398841 & 0.00363 & 0.3539 & $2f_{1}-f_{4}$ \\
$F12\textsubscript{Kepler}$ & 1.380730 & 0.00328 & 0.2701 & $f_{1}+f_{2}+f_{4}$ \\
\hline
\end{tabular}
\end{adjustbox}
\end{center}  
\end{table}

\begin{table}
\caption{Frequency solution for \textbf{KIC\,7304385} (ASASJ195052+4248.1).}
\label{fs13}
\begin{center}
\begin{adjustbox}{max width=0.5\textwidth}
\begin{tabular}{lllll}
\hline
\hline
(1) & (2) & (3) & (4) & (5) \\
Freq.No. & Frequency & Amplitude & Phase & ID \\
\hline
$F1\textsubscript{Kepler}$ & 1.269238 & 0.02562 & 0.9781 & $f_{1}$ \\
$F2\textsubscript{Kepler}$ & 1.418047 & 0.02075 & 0.1262 & $f_{2}$ \\
$F3\textsubscript{Kepler}$ & 1.462331 & 0.00796 & 0.6361 & $f_{3}$ \\
$F4\textsubscript{Kepler}$ & 0.148812 & 0.00671 & 0.0651 & $f_{2}-f_{1}$ \\	
$F5\textsubscript{Kepler}$ & 1.487632 & 0.00467 & 0.1299 & $f_{4}$ \\
$F6\textsubscript{Kepler}$ & 2.687285 & 0.00408 & 0.4037 & $f_{2}+f_{1}$ \\	
$F7\textsubscript{Kepler}$ & 1.120411 & 0.00296 & 0.9132 & $2f_{1}-f_{2}$ \\	
$F8\textsubscript{Kepler}$ & 1.243951 & 0.00301 & 0.9347 & $f_{1}+f_{3}-f_{4}$ \\
$F9\textsubscript{Kepler}$ & 2.538475 & 0.00294 & 0.2640 & $2f_{1}$	\\
$F10\textsubscript{Kepler}$ & 0.635646 & 0.00242 & 0.1740 & $f_{5}$ \\
$F11\textsubscript{Kepler}$ & 0.193106 & 0.00242 & 0.5330 & $-f_{1}+f_{3}$ \\	
$F12\textsubscript{Kepler}$ & 1.566863 & 0.00229 & 0.4698 & $2f_{2}-f_{1}$ \\	
$F13\textsubscript{Kepler}$ & 1.076156 & 0.00234 & 0.8462 & $2f_{1}-f_{3}$ \\	
$F14\textsubscript{Kepler}$ & 1.180661 & 0.00211 & 0.3085 & $f_{1}-2f_{3}+2f_{2}$ \\	
$F15\textsubscript{Kepler}$ & 0.152996 & 0.00184 & 0.8403 & $-2f_{1}+f_{2}+2f_{5}$ \\	
$F16\textsubscript{Kepler}$ & 2.053669 & 0.00175 & 0.5396 & $f_{2}+f_{5}$ \\	
$F17\textsubscript{Kepler}$ & 1.116238 & 0.00167 & 0.9054 & $f_{6}$ \\
$F18\textsubscript{Kepler}$ & 1.095130 & 0.00166 & 0.9118 & $2f_{1}+f_{3}-f_{2}-f_{4}$ \\	
$F19\textsubscript{Kepler}$ & 1.031884 & 0.00152 & 0.2022 & $2f_{1}+f_{2}-2f_{3}$ \\	
$F20\textsubscript{Kepler}$ & 1.378640 & 0.00152 & 0.8504 & $f_{1}+2f_{6}-f_{4}-f_{5}$ \\	
$F21\textsubscript{Kepler}$ & 2.836090 & 0.00150 & 0.6154 & $2f_{2}$ \\
\hline
\end{tabular}
\end{adjustbox}
\end{center}  
\end{table}

\begin{table}
\caption{Frequency solution for \textbf{HD\,211394} (BD-17\,6481).}
\label{fs14}
\begin{center}
\begin{adjustbox}{max width=0.5\textwidth}
\begin{tabular}{lllll}
\hline
\hline
(1) & (2) & (3) & (4) & (5) \\
Freq.No. & Frequency & Amplitude & Phase & ID \\
\hline
$F1\textsubscript{ASAS-3}$ & 0.452407	& 0.049	& 0.7275 & $f_{1}$ \\
$F2\textsubscript{ASAS-3}$ & 0.371387	& 0.027	& 0.6153 & $f_{2}$ \\
$F3\textsubscript{ASAS-3}$ & 0.904804	& 0.020	& 0.7111 & $2f_{1}$ \\
$F4\textsubscript{ASAS-3}$ & 0.823776	& 0.018	& 0.6548 & $f_{1}+f_{2}$ \\
$F5\textsubscript{ASAS-3}$ & 1.276179	& 0.014	& 0.6848 & $f_{2}+2f_{1}$  \\
\hline
$F1\textsubscript{ROAD}$ & 0.452625 & 0.051 & 0.0744 & $f_{1}$ \\
$F2\textsubscript{ROAD}$ & 0.370860 & 0.026 & 0.7140 & $f_{2}$ \\
$F3\textsubscript{ROAD}$ & 0.823041 & 0.020 & 0.4374 & $f_{1}+f_{2}$ \\
$F4\textsubscript{ROAD}$ & 0.904372 & 0.018 & 0.1150 & $2f_{1}$ \\
\hline
$F1\textsubscript{Kepler}$ & 0.452538 & 0.04193 & 0.8082 & $f_{1}$ \\
$F2\textsubscript{Kepler}$ & 0.372089 & 0.01975 & 0.7546 & $f_{2}$ \\
$F3\textsubscript{Kepler}$ & 0.905160 & 0.01774 & 0.2723 & $2f_{1}$ \\
$F4\textsubscript{Kepler}$ & 0.824686 & 0.01274 & 0.4111 & $f_{1}+f_{2}$ \\	
$F5\textsubscript{Kepler}$ & 0.080533 & 0.01063 & 0.3164 & $f_{1}-f_{2}$ \\
$F6\textsubscript{Kepler}$ & 1.005225 & 0.01002 & 0.3856 & $f_{3}$ \\
$F7\textsubscript{Kepler}$ & 0.533652 & 0.00904 & 0.2196 & $2f_{1}-f_{2}$ \\
$F8\textsubscript{Kepler}$ & 1.357718 & 0.00816 & 0.2133 & $f_{4}$ \\
\hline
\end{tabular}
\end{adjustbox}
\end{center}  
\end{table}             			

\begin{table}
\caption{Frequency solution for \textbf{GSC\,02780-02174} (TYC\,2780-2174-1).}
\label{fs15}
\begin{center}
\begin{adjustbox}{max width=0.5\textwidth}
\begin{tabular}{lllll}
\hline
\hline
(1) & (2) & (3) & (4) & (5) \\
Freq.No. & Frequency & Amplitude & Phase & ID \\
\hline
$F1\textsubscript{ASAS-SN}$ & 1.059750 & 0.034 & 0.3190 & $f_{1}$ \\
$F2\textsubscript{ASAS-SN}$ & 1.245288 & 0.013 & 0.8264 & $f_{2}$ \\
$F3\textsubscript{ASAS-SN}$ & 1.476331 & 0.008 & 0.9466 & $f_{3}$ \\
$F4\textsubscript{ASAS-SN}$ & 1.114903 & 0.008 & 0.4843 & $f_{4}$ \\
\hline
\end{tabular}
\end{adjustbox}
\end{center}  
\end{table}     
        	
\clearpage

%%%%%%%%%%%%%%%%%%%%%%%%%%%%%%%%%%%%%%%%%%%%%%%%%%%%%%%%%%%%%

\section{Fourier amplitude spectra} \label{fourier_spectra} 

This section provides the Fourier amplitude spectra of all HAGDOR stars, based on unwhitened data. The y-axes denote semi-amplitudes, as derived with \textsc{PERIOD04}. The employed data source is indicated in the panels. More information on the period analysis is provided in Section \ref{data_analysis}.

\begin{figure*}
\begin{center}
\includegraphics[width=1.00\textwidth]{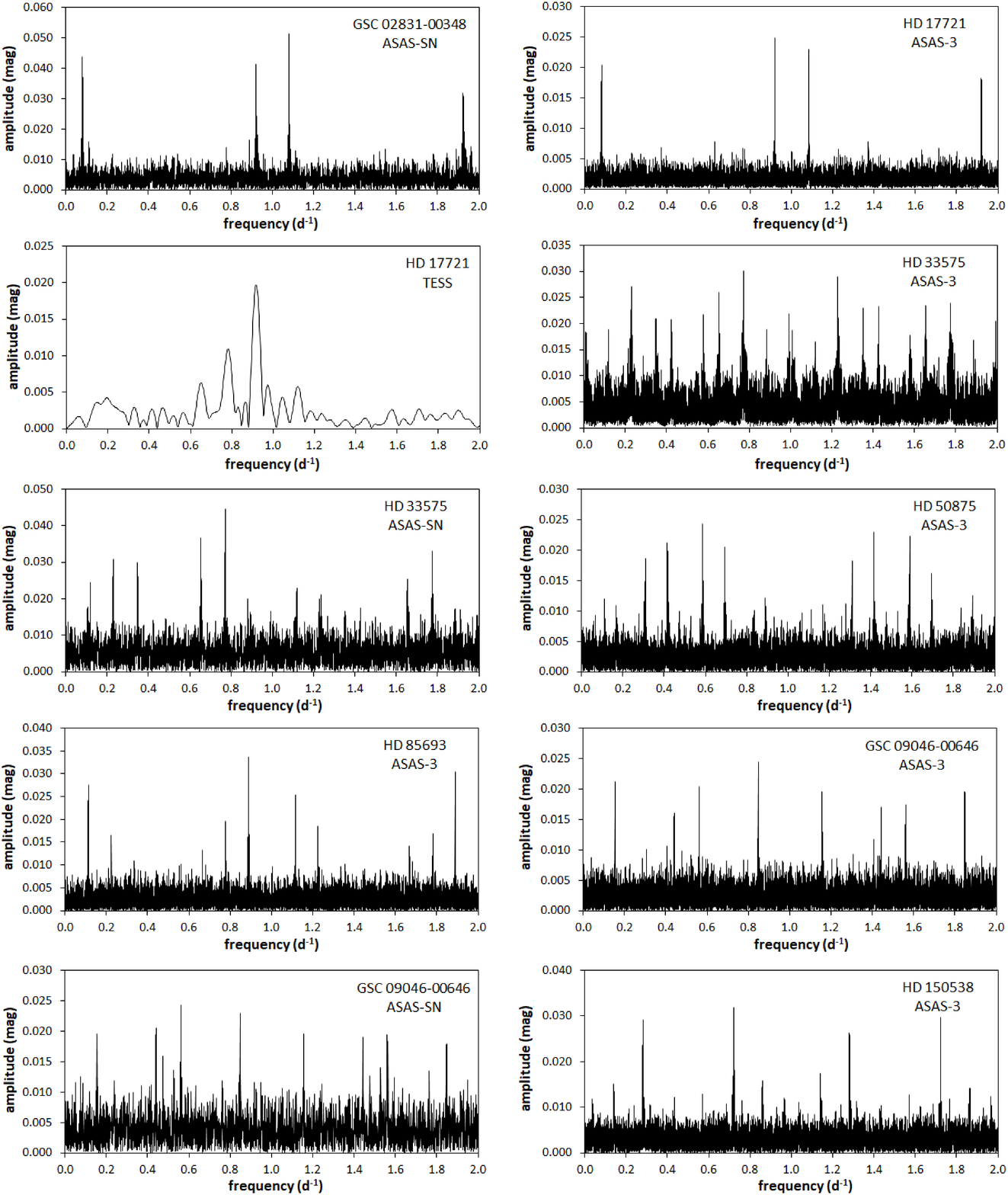}
\caption{Fourier amplitude spectra for all HAGDOR stars, based on unwhitened data. The employed data source is indicated in the panels.}
\label{fig_amplitude_spectra}
\end{center}
\end{figure*} 
\setcounter{figure}{0}
\begin{figure*}
\begin{center}
\includegraphics[width=1.00\textwidth]{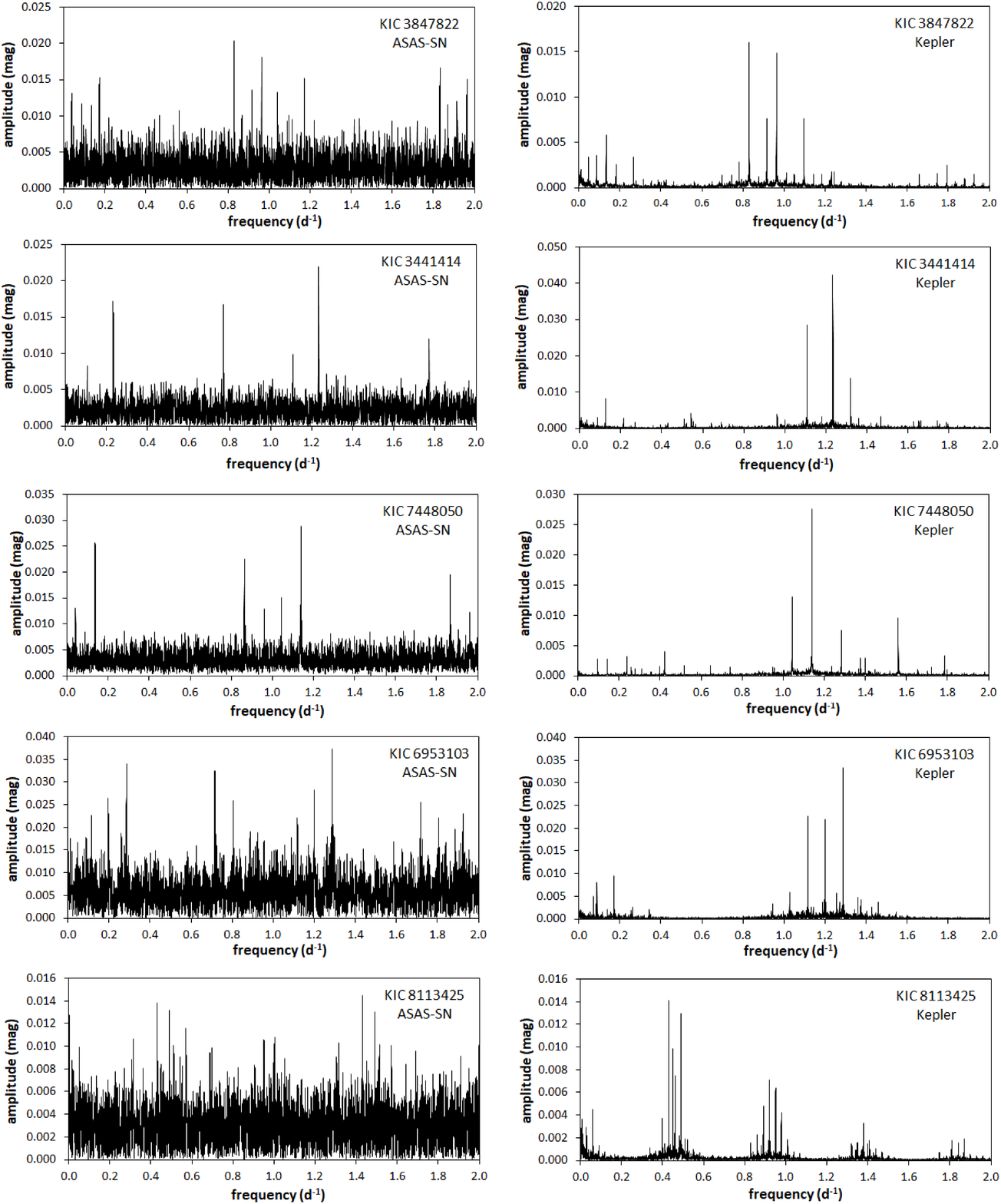}
\caption{Fourier amplitude spectra for all HAGDOR stars, based on unwhitened data. The employed data source is indicated in the panels.}
\label{fig_amplitude_spectra2}
\end{center}
\end{figure*}
\setcounter{figure}{0}
\begin{figure*}
\begin{center}
\includegraphics[width=1.00\textwidth]{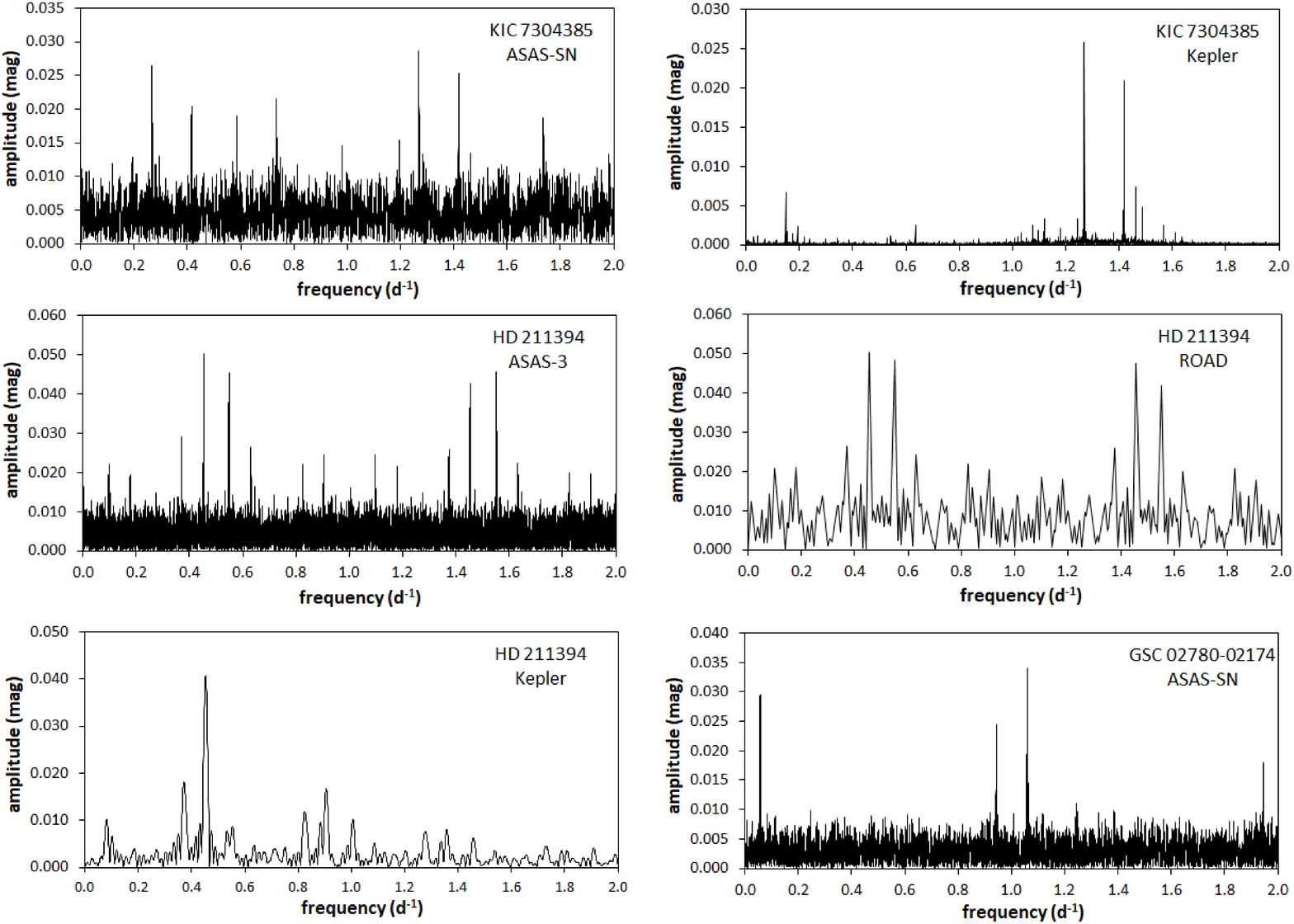}
\caption{Fourier amplitude spectra for all HAGDOR stars, based on unwhitened data. The employed data source is indicated in the panels.}
\label{fig_amplitude_spectra3}
\end{center}
\end{figure*} 

%\fi

% Don't change these lines
\bsp	% typesetting comment
\label{lastpage}
\end{document}